\documentclass[trackchanges]{aastex701}

\usepackage{multirow}
\usepackage{upgreek}
\usepackage{appendix}
\begin{document}

\title{Constraining turbulent solar flare acceleration regions by connecting kinetic modeling and X-ray observations}

\author[orcid=0000-0002-6060-8048,sname='Stores',gname='Morgan']{Morgan Stores}
\affiliation{University of Minnesota, School of Physics and Astronomy, Minneapolis, 55455, USA}
\affiliation{University Corporation of Atmospheric Research, Boulder, Colorado, 80301, USA}
\affiliation{Northumbria University, Newcastle upon Tyne, NE1 8ST, UK}
\email[show]{mstores@umn.edu}  

\author[orcid=0000-0001-6583-1989,sname='Jeffrey', gname='Natasha']{Natasha Jeffrey} 
\affiliation{Northumbria University, Newcastle upon Tyne, NE1 8ST, UK}
\email{natasha.jeffrey@northumbria.ac.uk}

\author[orcid=0000-0002-3992-8231, gname=Ewan,sname=Dickson]{Ewan Dickson}
\affiliation{Institute of Physics \& Kanzelhöhe Observatory, University of Graz, Universitätsplatz 5, 8010 Graz, Austria}
\email{fakeemail3@google.com}

\author[orcid=0000-0002-7863-624X,gname=James,sname=McLaughlin]{James McLaughlin}
\affiliation{Northumbria University, Newcastle upon Tyne, NE1 8ST, UK}
\email{fakeemail3@google.com}

\author[orcid=0000-0002-8078-0902,gname=Eduard,sname=Kontar]{Eduard Kontar}
\affiliation{University of Glasgow, Glasgow, G12 8QQ, UK}
\email{fakeemail3@google.com}

\begin{abstract}
Spatially-resolved X-ray observations are the key to understanding electron acceleration in solar flares. Currently, the underlying processes that efficiently energize solar flare particles are poorly constrained. Abundant flare observations suggest that turbulence plays a crucial role in transferring energy between the magnetic field and energetic electrons.
For the first time, we connect inhomogeneous acceleration from turbulence and hard X-ray spectroscopy and imaging observations with kinetic modeling to constrain the properties of flare acceleration. Observing three large flares with RHESSI, or Solar Orbiter/STIX, we extract X-ray imaging and spectroscopy observables. We compare with modeling results, mapping observables to electron acceleration and turbulent properties. We determine that extended regions of turbulence are required to match multiple X-ray observables, suggesting electrons are accelerated over a large fraction ($\sim 25\%$) of the flare loop; a property that is usually unconstrained from X-ray observations alone. Additionally, we determine acceleration timescales that vary between 7 and $22 \, \rm{s}$ by using fixed values for the turbulent scattering timescale and the velocity dependence of the acceleration diffusion coefficient. These fixed values are effectively unconstrained, but yield acceleration timescales that will help to restrict possible viable stochastic models.
\end{abstract}

\keywords{\uat{Solar Flares}{1496} --- \uat{Solar energetic particles}{1491} --- \uat{Solar physics}{1476} --- \uat{Solar x-ray flares}{1816}}

\section{Introduction} 

The solar corona is a hive of explosive activity. Solar flares are the result of distorted magnetic field lines reconnecting \citep{1958IAUS....6..123S,priest2002magnetic}, releasing up to $10^{32}$ ergs of energy and creating post-flare coronal loops, observed with soft X-ray (SXR) and extreme ultraviolet (EUV) emissions. Part of the energy released accelerates electrons out of the thermal distribution to energies $>20 \, \rm{keV}$ \citep{1969ApJ...158L.159F}. However, current understanding of the region accelerating electrons is limited, with competing acceleration mechanisms.

For several decades, the imaging spectroscopy abilities of X-ray instruments, e.g., the Reuven Ramaty High-Energy Solar Spectroscopic Imager \citep[RHESSI; ][]{2002SoPh..210....3L} and the Spectrometer Telescope for Imaging X-rays \citep[][ STIX]{2020A&A...642A..15K} onboard Solar Orbiter \citep[SolO;][]{2020A&A...642A...1M}, have provided spatially resolved X-ray spectra of coronal looptop and hard X-ray (HXR) footpoint sources \citep[e.g., ][]{2003ApJ...595L.107E,2022SoPh..297...93M}. These observations have validated the thick-target model \citep{1971SoPh...18..489B} in which electrons are accelerated in the corona, many of which then propagate down the legs of coronal loops and collide with the dense chromospheric plasma. Ensuing Coulomb collisions force the accelerated electrons to lose their energy, producing electron-ion bremsstrahlung X-ray emission \citep{holman2011implications} in the process. Additionally, recent models consider electron transport through a `warm-target' \citep{2015ApJ...809...35K,2019ApJ...871..225K} in which non-thermal electron properties are constrained through the coronal plasma properties. The warm-target model is not limited by the well recognized `low-energy cut off', a problem which burdens the cold thick-target model, thus allowing the power associated with non-thermal electrons to be constrained.
However, the exact mechanism behind this efficient electron acceleration \cite[reviews on electron acceleration in solar flares][]{1997JGR...10214631M, 2012SSRv..173..535P} is unconstrained and the lower density in the solar corona hinders direct observation of this region in X-rays. Thus, accelerated electron properties are usually inferred from the X-ray emission of the emitting electrons mainly emanating from the dense chromosphere. Additionally, dominant places of acceleration in solar flares are not well constrained. Observational evidence suggest possible acceleration regions may be in the coronal loop top or in current sheets \citep[e.g.][]{2016ApJ...828..103T} or in the above loop sources \citep{1994ApJS...90..623M,2012ApJ...748...33C}. In this study we focus on the loop top where we often see signs of acceleration \citep[e.g.][]{kontar2017turbulent,2021ApJ...923...40S,2024ApJ...973...96A}.

There are several proposed acceleration mechanisms, such as the formation of magnetic islands due to plasma instabilities \citep{2006Natur.443..553D,2011NatPh...7..539D,2014PhPl...21i2304D}, shock acceleration \citep[e.g.][]{2015Sci...350.1238C}, and magnetic energy dissipation by plasma waves and turbulence \citep{1993ApJ...418..912L,2012SSRv..173..535P,2017PhRvL.118o5101K}. For reviews on possible acceleration mechanisms see, \cite{1997JGR...10214631M,2009hppl.book.....S}. However, the exact mechanisms and locations of energy release and/or acceleration are not well-constrained.

In recent years, the dynamic nature of the flare environment has favored stochastic, second-order Fermi-type acceleration \citep{1949PhRv...75.1169F}. Stochastic acceleration \citep{1966PhRv..141..186S} provides enough energy to accelerate electrons to agree with X-ray observations \citep{2012SSRv..173..535P}. One possible mechanism capable of creating stochastic acceleration is magnetohydrodynamic (MHD) plasma turbulence \citep{1995ApJ...438..763G}. MHD plasma turbulence has been modeled extensively \citep{2023ApJ...944..147A} and is able to recreate observational evidence \citep{2023ApJ...947...67R} by transferring energy from large-to-small scales \citep{1993ApJ...418..912L}. 

Several have previously considered injecting a pre-accelerated electron distribution into the coronal loop in order to study electron transport effects \citep[for example,]{1982ApJ...259..341B,1990ApJ...359..524M,jeffrey2014spatial}. Alternatively, electron acceleration may be studied by including an escape term, such as in the leaky box model \citep{2013ApJ...777...33C}. However, these methods do not allow electron acceleration to occur at the same time as the transport effects. Stackhouse and Kontar 2018, allow for electron transport to occur simultaneously and introduced a spatially-dependent turbulent acceleration diffusion coefficient, highlighting the need to account for a spatial variation in turbulence. A key sign of plasma turbulence is non-thermal broadening of spectral lines, a feature frequently observed in flares \citep{2014ApJ...788...26D,2018ApJ...854..122W}.  
\cite{2021ApJ...923...40S} studied the non-thermal line broadening of three ions observed by the Hinode \citep{2007SoPh..243....3K} EUV imaging spectrometer (EIS) \citep{2007SoPh..243...19C} for one flare, and observed turbulence across the entire flare, both in the loop apex and loop legs.

Recently, \cite{2023ApJ...946...53S} studied how different properties of a turbulent acceleration region changed observed electron properties, using a kinetic transport model which included an extended turbulent acceleration region. The model produced outputs that can be directly compared to X-ray observations and several spectral and imaging diagnostics were created to constrain the acceleration region properties. 
Here, we obtain these spectral and imaging diagnostics using X-ray imaging and spectroscopy for three flares. The observed diagnostics are then compared to model outputs described in \cite{2023ApJ...946...53S} to constrain the properties of the acceleration region.

With regards to understanding solar flare electron acceleration, turbulence models are appealing, but have many uncertain parameters. We choose to employ turbulence in the acceleration region since many different mechanisms ultimately use the presence of turbulence to transfer energy to particle scales, and multiple solar flare observations \citep[e.g.,][]{1993ApJ...418..912L,2012SSRv..173..535P,2017PhRvL.118o5101K,2017SSRv..212.1107K} show the presence of turbulence, suggesting it does play a role in this energy transfer. Hence, constraining the properties of turbulence in the region is vital for all mechanisms, and the properties we constrain will help to ultimately constrain the mechanism itself. To comprehend the mechanisms and locations of particle acceleration, it is essential to connect i) modeling of turbulent acceleration of electrons and their subsequent transport at the Sun, and ii) diagnostics of electron acceleration and transport from multiple HXR observations. Here, for the first time, we combine i) and ii) to partially constrain the properties of the turbulent acceleration region. 
\S \ref{sect:obs} provides an overview of the flares in this study and the imaging and spectroscopy analysis, 
\S \ref{sect:kinetic} outlines the kinetic model, in \S \ref{sect:results} we discuss the main results of the study, and \S \ref{sect:discussion} summarizes the study and considers possible mechanisms creating the turbulence.

\section{Observational Overview}\label{sect:obs}

Here, we study three flares during their impulsive phase in order to constrain acceleration region properties using the X-ray diagnostics presented in \cite{2023ApJ...946...53S}. In order to measure the spectral and imaging diagnostics, all three flares, when observed by the relevant X-ray instrument, are situated on the solar limb with at least one HXR chromospheric footpoint visible to the observing instrument. This allows the coronal loop length to be determined and gives a clear separation between the coronal looptop and footpoint X-ray sources. Additionally, each flare produced non-thermal emission above $50 \, \rm{keV}$ which is required to determine several spectral diagnostics.  

An overview of the flares studied is \S \ref{sect:events} and details of the imaging and spectral observations for each flare is given in \S \ref{Observing: Imaging Diagnostics} and \S \ref{sect:spectro_diagnostics}, respectively. Imaging-spectroscopy diagnostics are discussed in \S \ref{sect:chap3:imaging_spect}.

\subsection{Event Summary}\label{sect:events}

We study three flares during their impulsive phase; whilst we do not perform a statistical study, this paper aims to compare the results of multiple events. Two of the flares were observed by RHESSI: 
SOL2011-02-24T07:35 (Flare 1) a M3.5 class flare observed on the 24th February 2011 at 07:29:25 UT, and SOL2013-05-13T01:53 (Flare 2) a X1.7 class flare observed on the 13th May 2013 at 02:07:55 UT. The third flare was observed by SolO/STIX:
SOL2022-03-28T10:58 (Flare 3), a M4.1 class flare on the 28th March 2022 at 11:15:28 UT \footnote{Earth orbiting satellites observed the emission from Flare 3 at 11:21:03.7 UT + LT, where LT = $335.7 \, \rm{s}$ is the light travel time to Earth for this observation.}. 
Table \ref{tab:flare_times} lists the studied flares, the observing X-ray instrument, the Geostationary Operational Environmental Satellite \citep[GOES; ][]{1991SoPh..133..371K} class, and the date and time of the observations used in this study. 

\begin{figure*}[t!]
    \centering
    \includegraphics[width = \linewidth]{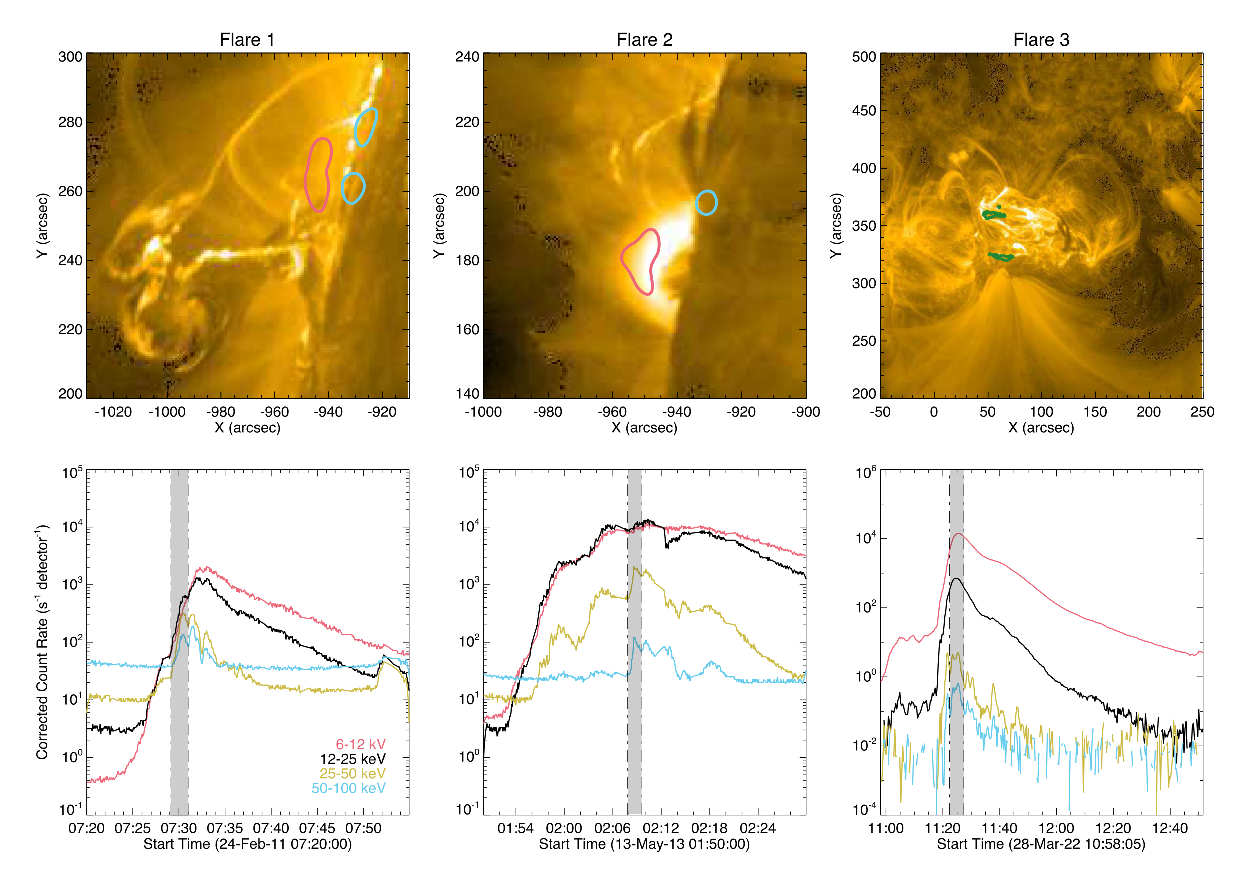}
    \caption{\textit{Top row}: SDO/AIA $171$ \AA \, images of Flare 1 (left), Flare 2 (middle), and Flare 3 (right), respectively. Pink and blue contours show the 50\% intensity levels of the  $6-12 \, \rm{keV}$ and $50-100 \, \rm{keV}$ X-ray emission, respectively. Flare ribbons observed in $1700$ \AA \, are shown by the green $50\%$ contours for Flare 3. \textit{Bottom Row}: Corresponding light curves for each event. The grey shaded boxes highlight the time periods studied here for each flare.}
    \label{fig:context}
\end{figure*}

\begin{figure}[t!]
    \centering
    \includegraphics[width=0.8\linewidth]{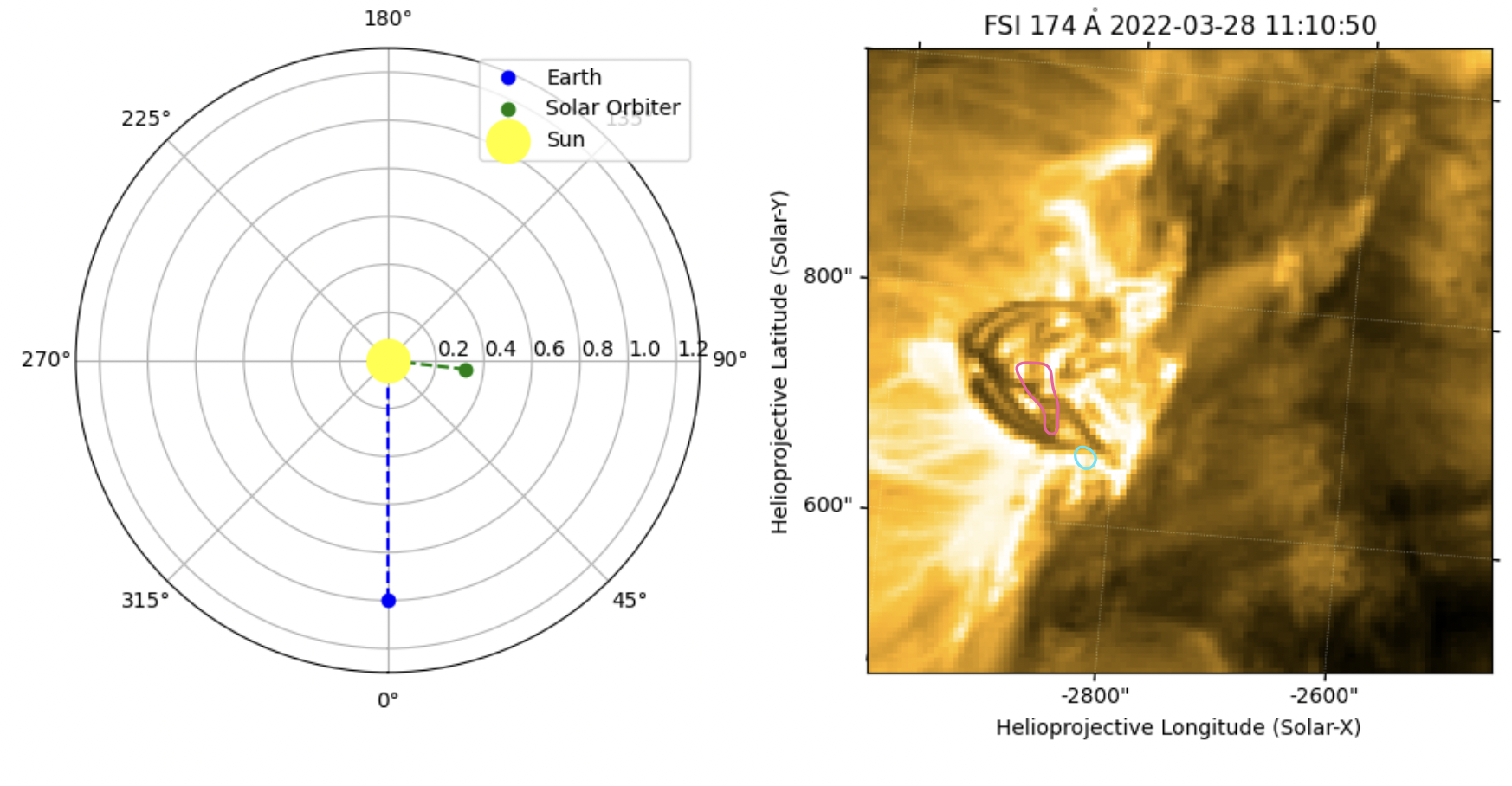}
    \caption{Left: SolO/STIX location during Flare 3 with respect to Earth and the Sun. The angle of separation between SolO and Earth was 83.4 degrees at the time of observation. Right: FSI/SolO 174\AA image of Flare 3.}
    \label{fig:Solo_view}
\end{figure}

Images by the Solar Dynamics Observatory's (SDO) Atmospheric Imaging Assembly (AIA) \citep{2012SoPh..275...17L} provide context to these observations in Figure \ref{fig:context}. RHESSI $50\%$ contours of the $6-12 \, \rm{keV}$ and $50-100\,  \rm{keV}$ emissions are shown in pink and blue, respectively, for Flares 1 and 2. For Flare 3, $50\%$ contours in green highlight the ribbons observed in AIA 1700 \AA \footnote{For consistency here, all three flares are shown from Earth's viewpoint. The angle of separation between SolO/STIX and Earth is $83.4^{\circ}$ at the time of observation, with STIX observing the flare on the eastern limb, see Figure \ref{fig:Solo_view}}. The bottom row of Figure \ref{fig:context} displays light curves from RHESSI (Flares 1 and 2) and STIX (Flare 3) showing the HXR emission, corresponding to accelerated electrons where observed times are highlighted by the gray boxes. All three flares were studied for $\approx 100$ seconds, covering the first peak in their HXR emission \textit{which included a noticeable increase in emission above 50 keV}. It should be noted that Flare 2 sees a small peak in $25-50 \, \rm{keV}$ emission at an earlier time, which is further discussed in \S \ref{sect:spectro_diagnostics}. Additionally, Flare 3 has a very short peak, just prior to the observation time in this study; this peak was not selected due to its very short duration.

 \begin{table*}[t!]
   \small
    \centering
    \caption{Overview of flares studied: flare ID, the observing X-ray instrument, the GOES class, date of observation and the time of observation used in this study.}
    \begin{tabular}{c c c c c c}
    \hline 
    \hline 
    & Flare ID & Observing & GOES  & Date & Observation time [UT] \\
    & & Instrument & Class & & \\
    \hline 
        Flare 1  & SOL2011-02-24T07:35& RHESSI & M3.5 & 24 February 2011 & 07:29:25 - 07:31:05 \\
        Flare 2  & SOL2013-05-13T01:53& RHESSI & X1.7 & 13 May 2013 & 02:07:55 - 02:09:35 \\
        Flare 3  & SOL2022-03-28T10:58& SolO/STIX & M4.1 & 28 March 2022 & 11:15:28 - 11:16:44\\
        \hline 
    \end{tabular}
    \label{tab:flare_times}
\end{table*}
\normalsize

All three flares have been previously studied and several of the studies have provided evidence of magnetic reconnection, non-thermal X-ray sources, and/or electron acceleration. In Flare 1 \citep[e.g.,][]{2011A&A...533L...2B, 2012ApJ...753L..26M}, higher electron rates were observed in the coronal looptop source compared to the footpoints \cite{2013A&A...551A.135S}, concluding that electrons were trapped in the corona, possibly indicating the presence of coronal turbulence. Flare 2 was observed by several instruments as discussed in \cite{2017ApJ...835...43S,2023ApJ...950...71S} using SXR, EUV and microwave emissions. Studies observe a `soft-hard-harder' spectrum, suggesting that electrons trapped in the loop are continuously accelerated. Additionally, \cite{2023ApJ...954...58S} considered Fermi first-order acceleration mechanisms to model the temporal and spatial HXR changes observed in this flare. 
Recently, the evolution of non-thermal sources in Flare 3 was studied by \cite{2023A&A...679A..99P}, determining at later times in the flare the location of the southern footpoint changed, possibly suggesting reconnection propagating along the arcade. Additionally, Flare 3 lies within a highly active region which produced four coronal mass ejections. Although we currently do not have direct evidence of coronal turbulence in this flare (this flare lacks Hinode EIS data), due to the strong evidence of turbulence in Flare 1 and 2, this flare was included in this study.

Unlike RHESSI\footnote{RHESSI sits at 1 AU with an Earth-centered orbit.}, SolO has an elliptical orbit which reaches a minimum of $\sim 0.3 \, \rm{AU}$ from the Sun. At the time of observation of Flare 3, SolO was situated at $0.325 \, \rm{AU}$ from the Sun with an angle of 83.4 degrees between the spacecraft and Earth, as shown in the left panel of Figure \ref{fig:Solo_view}. A 174\AA \, image of Flare 3 taken by SolO Full Sun Imager (FSI; \cite{2020A&A...642A...8R}) is shown in the right panel of Figure \ref{fig:Solo_view}, with the $50\%$ contours of $6-12 \, \rm{keV}$ and $36-74 \, \rm{keV}$ emissions shown in pink and blue, respectively. Due to the close proximity to the Sun, and the off-limb location of the event (when observed by SolO, see Figure \ref{fig:Solo_view}), Flare 3 is partially obscured from the full field of view of STIX leaving the top row of pixels of each detector in shadow. Thus, only the bottom row of pixels is considered in this analysis.

\subsection{Observing: Imaging Diagnostics}\label{Observing: Imaging Diagnostics}
\begin{figure*}[t!]
    \centering
    \includegraphics[width = 0.85\textwidth]{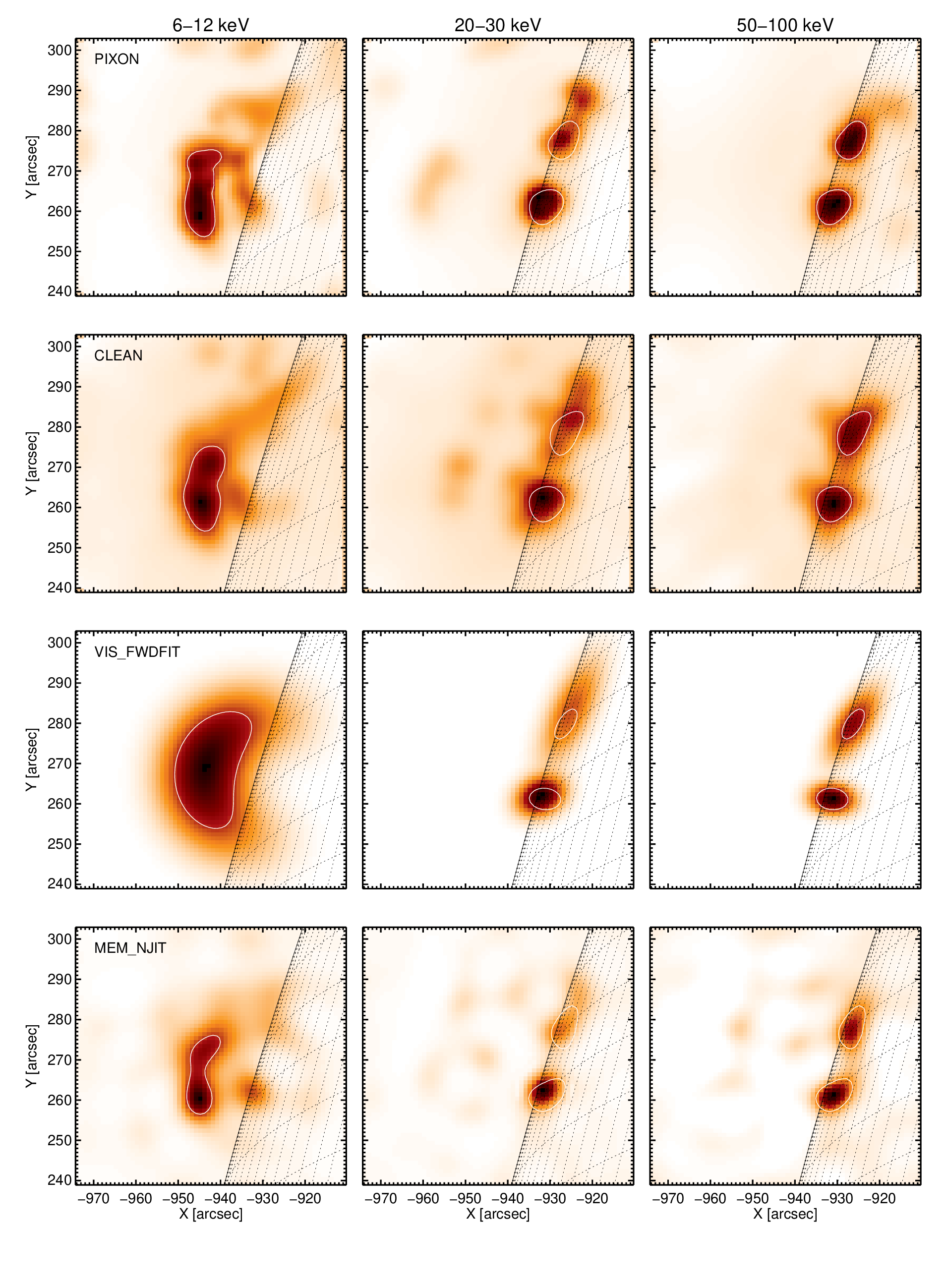}
    \caption{Flare 1 RHESSI imaging. Columns left to right: RHESSI $6-12 \, \rm{keV}$, $20-30 \, \rm{keV}$, and $50-100 \, \rm{keV}$ images. Rows show different imaging algorithms, top to bottom: Pixon, CLEAN (CLEAN beam width =  2.2), VIS\_FWDFIT, MEM\_NJIT. Column 1 shows a white contour of the looptop, defined as the  $50\%$ contour of the $6-12 \,\rm{keV}$ emission. The white contours in Columns 2 and 3 highlight footpoints, defined as the $50\%$ contour of the $50-100 \, \rm{keV}$ emission.}
    \label{fig:Feb_algorithm_comparasion}
\end{figure*}

\begin{figure*}[t!]
    \centering
    \includegraphics[width = 0.85\textwidth]{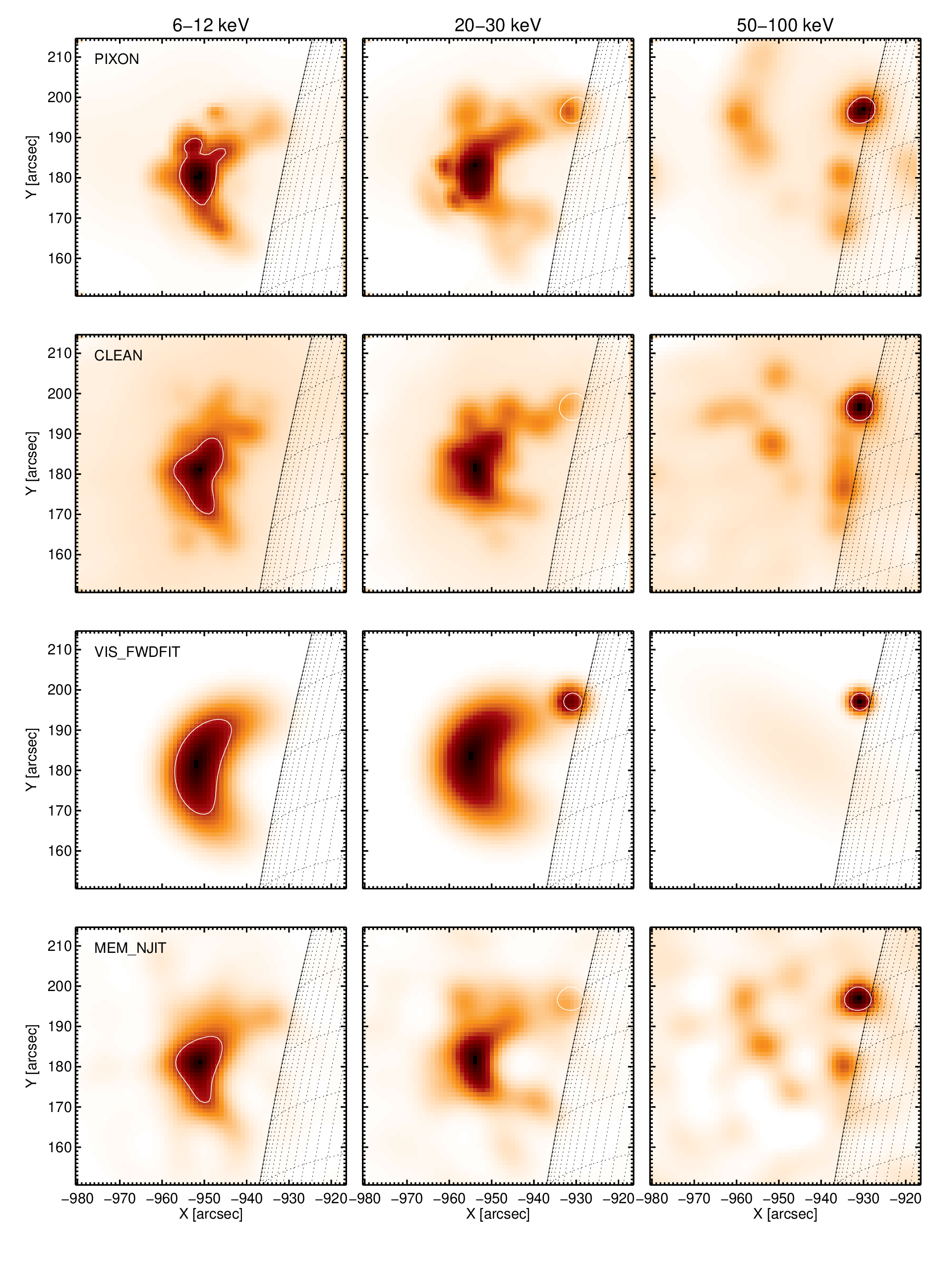}
    \caption{Flare 2 RHESSI imaging. Columns left to right: RHESSI $6-12 \, \rm{keV}$, $20-30 \, \rm{keV}$, and $50-100 \, \rm{keV}$ images. Rows show different imaging algorithms, top to bottom: Pixon, CLEAN (CLEAN beam width =  2.5), VIS\_FWDFIT, MEM\_NJIT. Column 1 shows a white contour of the looptop, defined as the $50\%$ contour of the $6-12 \,\rm{keV}$ emission. The white contours in Columns 2 and 3 highlight footpoints, defined as the $50\%$ contour of the $50-100 \, \rm{keV}$ emission.}
    \label{fig:May_algorithm_comparasion}
\end{figure*}

\begin{figure*}[t!]
    \centering
    \includegraphics[width = 0.85\textwidth]{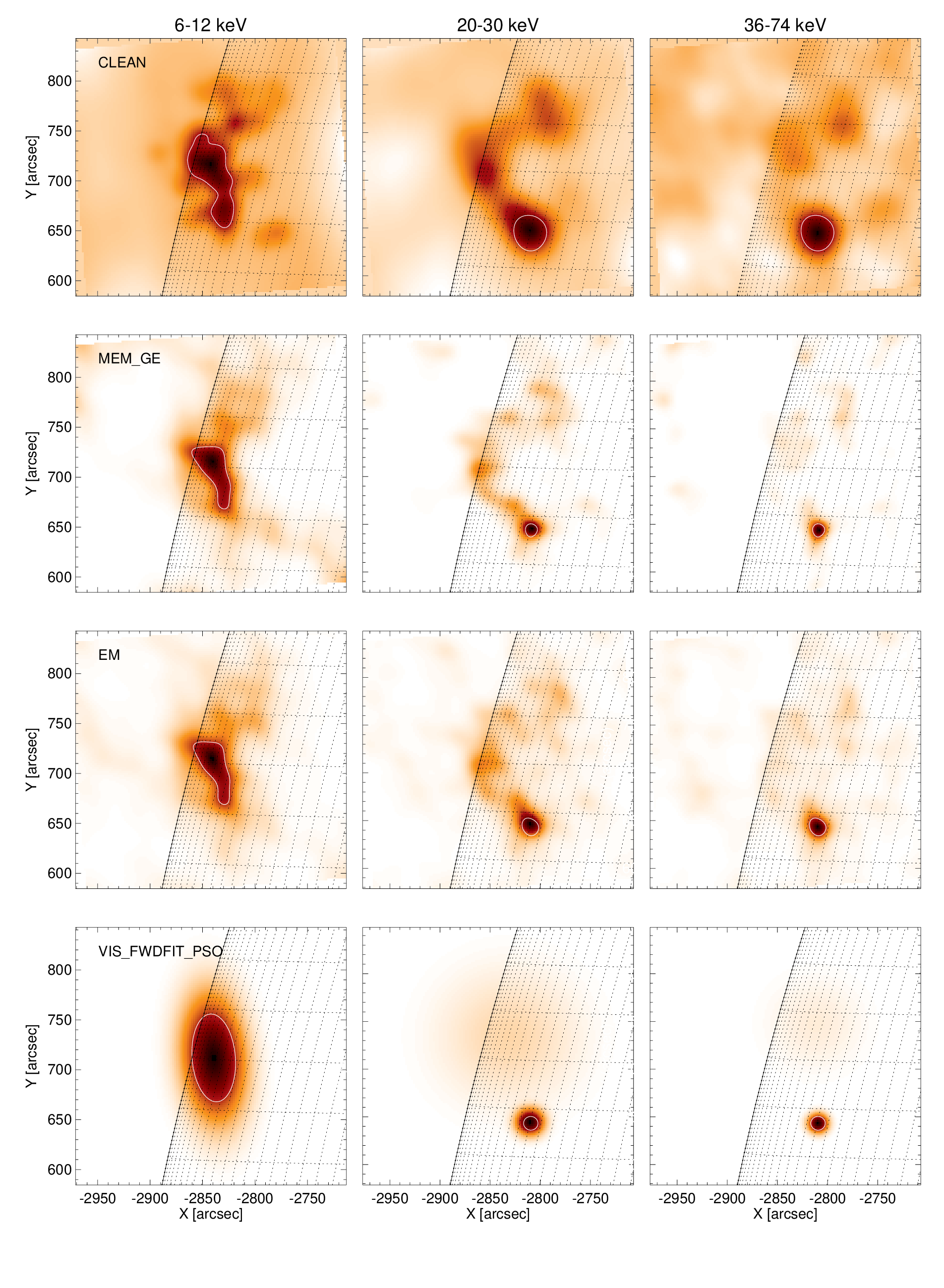}
    \caption{Flare 3, STIX imaging. Columns left to right: STIX $6-12 \, \rm{keV}$, $20-30 \, \rm{keV}$, and $36-74 \, \rm{keV}$ images. Rows show different imaging algorithms, top to bottom: CLEAN (CLEAN beam width =  2.0), MEM\_GE, EM, VIS\_FWDFIT\_PSO. Column 1 shows a white contour of the looptop, defined as the  $50\%$ contour of the $6-12 \,\rm{keV}$ emission. The white contours in Columns 2 and 3 highlight footpoints, defined as the $50\%$ contour of the $36-74 \, \rm{keV}$ emission.}
    \label{fig:march_algorithm_comparasion}
\end{figure*}

The shape of the X-ray sources was determined by comparing several imaging algorithms. For events observed by RHESSI (Flare 1 and Flare 2), the Pixon, CLEAN, VIS\_FWDFIT, and MEM\_NJIT imaging algorithms \citep{2002SoPh..210...61H} were compared, see Figures \ref{fig:Feb_algorithm_comparasion} and \ref{fig:May_algorithm_comparasion}. Images for Flare 3, observed by STIX, were created using the CLEAN, MEM\_GE, EM and VIS\_FWDFIT\_PSO \citep{2022A&A...668A.145V} imaging algorithms, see Figure \ref{fig:march_algorithm_comparasion}. 

For each flare studied, the looptop region is defined as the area covered by the $50\%$ contour of the $6-12 \,\rm{keV}$ energy range, highlighted by the white contour in the first column of Figures \ref{fig:Feb_algorithm_comparasion} - \ref{fig:march_algorithm_comparasion}. Similarly, the footpoint region is defined as the area covered by the $50\%$ contour of high energy non-thermal emission, for Flares 1 and 2 this is described by the $50-100 \,\rm{keV}$ energy range, shown in the final two columns in Figures \ref{fig:Feb_algorithm_comparasion} and \ref{fig:May_algorithm_comparasion}. For Flare 3, due to a lack of emission above $74 \, \rm{keV}$ the high non-thermal emission is described by the $36-74\, \rm{keV}$ energy range, as seen in the final two columns in Figure \ref{fig:march_algorithm_comparasion}. In the X-ray images, Flares 1 and 3 have one footpoint visible and Flare 2 has two footpoints. All three flares were observed by their respective X-ray instrument in order to get a clear separation between the coronal and chromospheric X-ray sources and to minimize projection effects.

For the CLEAN imaging algorithm, a CLEAN beam width of 2.2 and 2.5 were chosen for Flare 1 and Flare 2, respectively. These beam widths were chosen by comparing the size of the CLEAN X-ray sources to the sources made by other algorithms. 
For both events the Pixon, CLEAN, and MEM\_NJIT algorithms are in agreement about the size and shape of the X-ray sources, for all energy ranges studied. 
When creating the VIS\_FWDFIT images, for Flare 1, a loop model was used for the $6-12 \, \rm{keV}$ energy range and an elliptical Gaussian model was used to create the $20-30 \, \rm{keV}$ and $50-100 \, \rm{keV}$ energy ranges. For Flare 2, a loop model was successfully used in both the $6-12 \, \rm{keV}$ and the $20-30 \, \rm{keV}$ images. Additionally, a circular Gaussian model was used in the $20-30 \, \rm{keV}$ (alongside the loop model) and $50-100 \, \rm{keV}$ images.
Several previous studies such as \cite{2013ApJ...766...75J} determined coronal source properties using the VIS\_FWDFIT algorithm. However, the shape fitted by the VIS\_FWDFIT algorithm produced a larger source compared to the other algorithms, and thus, this study did not consider the VIS\_FWDFIT imaging algorithm.
For Flare 3, the VIS\_FWDFIT\_PSO algorithm was not successful in fitting the sources, creating a shape that was much larger than the other imaging algorithms. An elliptical Gaussian model was used to create the $6-12 \, \rm{keV}$ image and a circular Gaussian model for the $20-30 \, \rm{keV}$ and $50-100 \, \rm{keV}$ images. Furthermore, the CLEAN imaging algorithm did not match the size or shape of the sources created by the MEM\_GE and EM algorithms. Thus, neither CLEAN or VIS\_FWDFIT\_PSO was considered in the analysis of Flare 3. 

\begin{figure*}[t!]
    \centering
    \includegraphics[width = 0.9\textwidth]{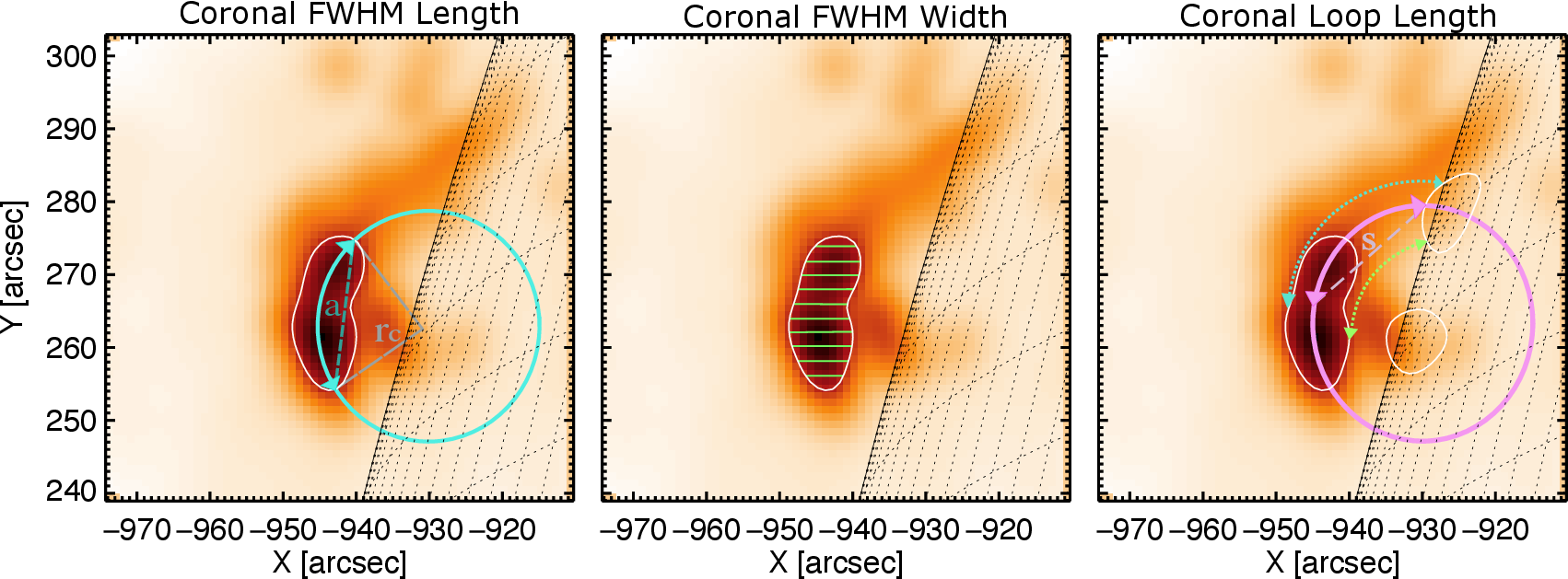}
    \caption{Example of how coronal source dimensions were determined from imaging. All panels show the $6-12 \, \rm{keV}$ CLEAN image for Flare 1. \textit{Left}: the length of the coronal source was determined, by fitting a circle of radius r to the source and calculating the length of the arc between the contours, indicated with the arrows, using the chord of length a. \textit{Middle}: measuring the coronal source FWHM width at several points to determine the average width. \textit{Right}: the length of the coronal loop being measured from looptop to footpoint, as indicated by the pink arrow, using the length of chord $S$. Upper (blue arrow) and lower (green arrow) coronal loop lengths were determined by fitting a circle to the outer and inner loop, respectively. This panel includes contours of the footpoint emission in white.}
    \label{fig:measuring}
\end{figure*}

The size of coronal features, i.e., coronal source full width at half maximum (FWHM) length and width, and the coronal loop length from loop apex to footpoint, were measured for all relevant imaging algorithms (the imaging algorithms included in this study) and an average value was taken for each flare. Figure \ref{fig:measuring} shows an example of how the coronal source FWHM length and width, and the coronal loop length were measured. The values of the imaging parameters $\pm$ uncertainties are given in Table \ref{chap4:tab:imaging_params}. 

The example given uses the CLEAN image created for Flare 1. Firstly, the left panel of Figure \ref{fig:measuring} depicts measuring the coronal source FWHM length $l$ [Mm], which is determined by fitting a circle to the $50\%$ contour of the $6-12\, \rm{keV}$ images to calculate the length of the arc, where $r_c$ [Mm] is the radius of the circle fitted, and $a$ [Mm] is a chord connecting the two ends of the coronal source FWHM. The coronal source FWHM length was then calculated using, 
\begin{equation}
	 \ell = \frac{2\pi r_c}{180}\rm{sin}^{-1}\left( \frac{\textit{a}}{2\textit{r}_\textit{c}} \right )
\end{equation}
The coronal source loop FWHM width $w$ [Mm] was determined by measuring the width of the $50\%$ contour in the 6-12 keV images in several locations (shown by the green lines in the middle panel of Figure \ref{fig:measuring}) and taking an average across the three imaging algorithms. The error on both $l$ and $w$ was calculated by repeating this process for the $45\%$ and $55\%$ contour regions. As in \cite{2019ApJ...871..225K}, assuming a cylindrical coronal loop, the volume of the coronal source, was estimated using 
\begin{equation}
V = \frac{\pi w^2 \ell }{ 4} \, . 
\end{equation}\label{eq:volume}
Finally, the length $L$ [Mm] from the coronal looptop to the chromospheric footpoint was determined by measuring the length of the arc connecting the coronal source to the footpoint, as shown in the right panel of Figure \ref{fig:measuring}. In Figure \ref{fig:measuring}, $S$ [Mm] is the length of the chord connecting the looptop and footpoint. The uncertainty was determined by replicating the fitting along the outer and inner edge of the coronal loop, indicated with the dashed lines in Figure \ref{fig:measuring}. 

\begin{table*}[t!]
    \centering
    \caption{Parameters obtained from imaging: length from looptop to footpoint $L$, coronal source length $l$, coronal source width $w$, coronal source volume $V$, coronal density $n$.}
    \begin{tabular}{c c c c c c}
    \hline 
    \hline
    & $L$ & $\ell$ & $w$  & $V$ & $n$ \\
    & [Mm] & [Mm] & [Mm] & [$\times 10 ^{27}$ cm$^3$] & [$\times 10 ^{10}$ cm$^{-3}$]  \\
    \hline 
        Flare 1  & $16\pm4$ & $15.4^{+0.8}_{-0.3}$ & $4.4\pm0.3$ & $0.23^{+0.04}_{-0.03}$ & $6\pm1$\\
        Flare 2  & $18^{+3}_{-4}$ & $14 \pm 1$ & $5.1\pm0.7$ & $0.28^{+0.09}_{-0.10}$ & $36^{+7}_{-8}$  \\
        Flare 3  & $24^{+13}_{-14}$ & $15\pm1$ & $5.1\pm0.6$ & $0.30^{+0.09}_{-0.08}$ & $6\pm1$ \\
        \hline 
    \end{tabular}
    \label{chap4:tab:imaging_params}
\end{table*}

\subsection{Observing: Spectroscopy Diagnostics}\label{sect:spectro_diagnostics}
Spatially-integrated X-ray spectroscopy with RHESSI and STIX provided values for several non-thermal electron and plasma properties, including the non-thermal electron power law spectral index, the coronal temperature $T$ [K], emission measure EM [cm$^{-3}$], total non-thermal electron flux $\dot{N}$ [electrons s$^{-1}$], and plasma density $n$ [cm$^{-3}$], where 
\begin{equation}
n = \sqrt{\frac{\rm{EM}}{Vf}} \, , 
\label{eq:density_emisison_measure}
\end{equation}
where $f$ is the filling factor (assumed to be unity, i.e $f = 1$) and assuming the plasma is isothermal with constant density. Although studies, such as \cite{jeffrey2015high}, show that coronal X-ray sources are multi-thermal and have strong vertical temperature and density gradients with a broad differential emission measure (DEM), an isothermal plasma with constant density is suitable approximation for comparison with the diagnostics shown in \cite{2023ApJ...946...53S}. To extract the properties, each flare was fitted using the following models and corresponding Object Spectral Executive \citep{2002SoPh..210..165S, 2020ascl.soft07018T} (OSPEX) fitting functions:
\begin{itemize}
\item Isothermal plus cold thick-target model (\textit{f\_vth}+\textit{f\_thick2}); 
\item Isothermal plus warm-target model (\textit{f\_vth}+\textit{f\_thick\_warm});
\item Warm-target model only (\textit{f\_thick\_warm}).
\end{itemize}
These models assume the plasma is isothermal with a constant density, as in \cite{2023ApJ...946...53S}. Previous studies show the cold target model is unable to constrain the so-called `low-energy cutoff' \citep{2013ApJ...769...89I,2019ApJ...871..225K}, a parameter required to constrain the power associated with non-thermal electrons. However, the warm-target model constrains this property by taking the background plasma properties ($T$, $n$, $L$) into account \citep{2019ApJ...871..225K}.

\begin{figure*}[t!]
    \centering
    \includegraphics[width = \linewidth]{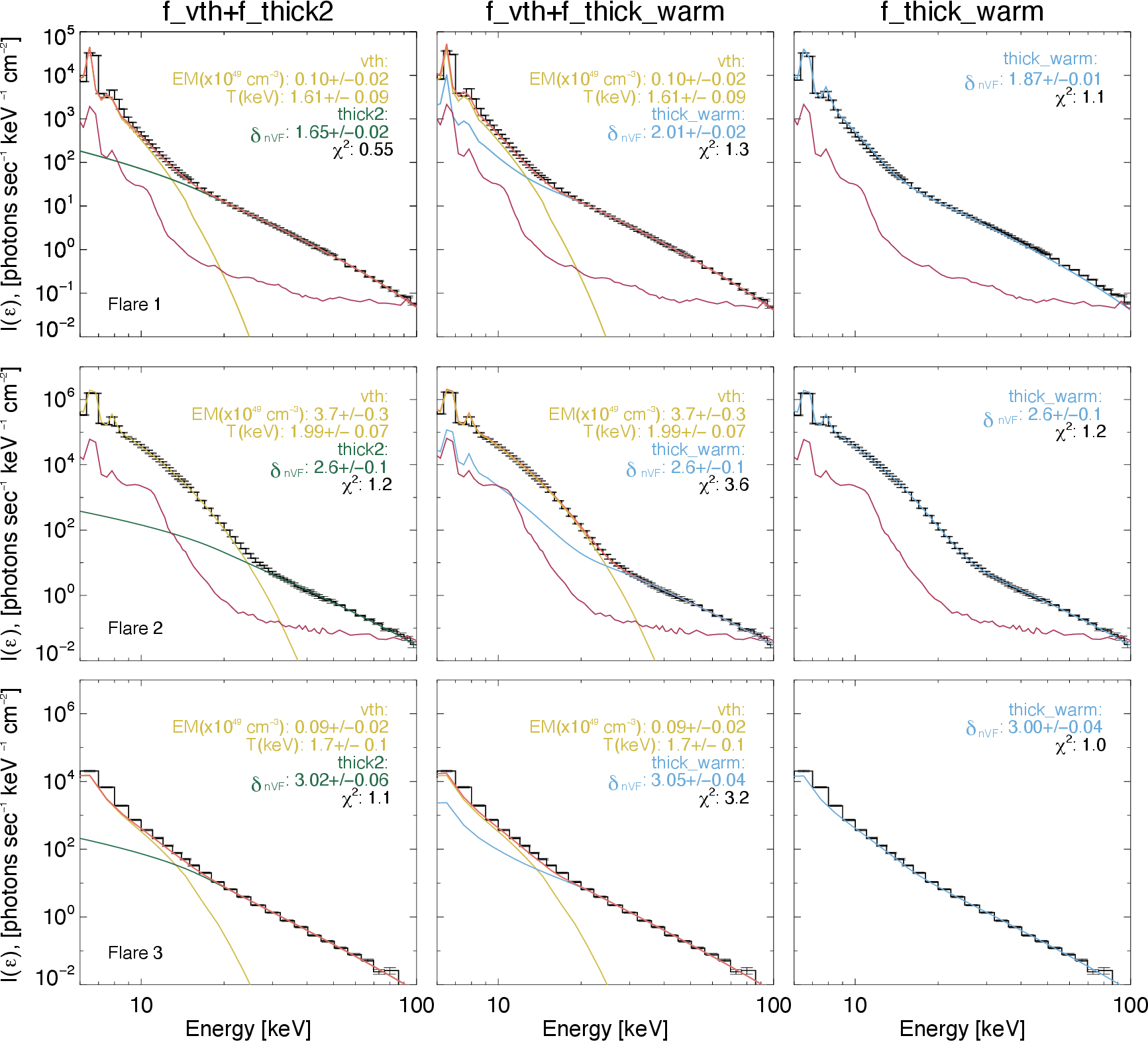}
    \caption{A comparison of spectral fittings for each flare. The left, middle and right columns show spectral fittings when using $f\_vth+f\_thick2$, $f\_vth+f\_thick\_warm$, and $f\_thick\_warm$, respectively. }
    \label{fig:spectra_comparasion}
\end{figure*}

Figure \ref{fig:spectra_comparasion} compares these three fitting models for each flare. The observed photon flux, $I$ [photons $\rm{s}^{-1} \, \rm{keV}^{-1}  \, \rm{cm}^{-2}$] energy spectrum is shown in black, $f\_vth$ in yellow,  $f\_thick2$ in green, $f\_thick\_warm$ in blue, and the total fit is in red. The background emission is shown in purple for the two RHESSI flares. However, (after private communication with the STIX team) at this time obtaining the STIX background data in photon flux is not possible, as such the background emission is absent from Figure \ref{fig:spectra_comparasion}. Gaussian lines (not shown in Figure \ref{fig:spectra_comparasion}) were also added to account for `bumps' in the spectrum below 10~keV, possibly due to iron and nickel lines that appear close to 6.7~keV and 8.1~keV.
An albedo component was also fitted to account for any X-ray Compton backscattering in the photosphere, which may account for up to 40\% of emission between $25-50 \, \rm{keV}$ \citep{1978ApJ...219..705B,2006A&A...446.1157K}. However, as the flares studied are off limb, there was not a significant change to the spectral fittings \citep{2011A&A...536A..93J}, i.e., the spectral index changed by $<1\%$ for Flare 1, $1\%$ for Flare 2, and $3\%$ for Flare 3. Thus, the final fittings do not contain an albedo component. 

To make the energy spectra, detectors [2, 4, 5, 6, 7, 8, 9] were combined for Flare 1 as the remaining detectors were not consistent at low energies. Similarly, only detector 5 was used for Flare 2. Due to the large size of Flare 2 (GOES class X1.7) the counts in detector 5 alone were sufficient to create the energy spectrum. All detectors were used for Flare 3. However, as previously mentioned, Flare 3 was partially obscured from the STIX field of view. Thus, only the bottom row of pixels was considered for each detector.

The value of emission measure and plasma temperature from the \textit{f\_vth}+\textit{f\_thick2} fitting was then used in the \textit{f\_vth}+\textit{f\_thick\_warm} fitting, following the method used in \citep{2019ApJ...871..225K}. Thus the emission measure given in Table \ref{Table_plasma_properties} is the value obtained from the \textit{f\_vth}+\textit{f\_thick2} fitting. However, the plasma temperature for each flare in Table \ref{Table_plasma_properties} is an average of the value obtained from \textit{f\_vth}+\textit{f\_thick2} and \textit{f\_thick\_warm} fittings, these values were similar resulting in an uncertainty of $\pm 1 \, \rm{MK}$ for each flare. 

Furthermore, the three spectral fitting models produced very similar values of the spectral index for each flare. \cite{2023ApJ...946...53S} discussed the spectral index of the density-weighted emitting electron distribution $\delta_{nVF}$, such that in a simple thick-target model, we would expect the inferred spectral index of an injected electron distribution $\delta \approx \delta_{nVF}+2$ and the X-ray photon index $\gamma \approx \delta_{nVF} + 1$. For easy comparison with model outputs, the given accelerated electron spectral index $\delta$ is converted to $\delta_{nVF}$. The value of $\delta_{nVF}$ given in Table \ref{Table_plasma_properties} is the average of the spectral fittings shown in Figure \ref{fig:spectra_comparasion} for each flare. The resulting error on these values is small; $\delta_{nVF} = 1.84\pm 0.03$, $2.6\pm 0.1 $, and $3.0\pm 0.1$, for Flares 1, 2, and 3, respectively.  

The light curve for Flare 2 (see Figure \ref{fig:context}) shows a peak in the $25-50 \, \rm{keV}$ emission at approximately 02:00 UT, around eight minutes prior to the observation time used in this study. This earlier time did not show a significant increase in $50-100 \, \rm{keV}$ emission, unlike the observed time beginning at 02:07:55 UT. The initial peak showed a much softer energy spectrum ($\delta_{nVF} = 4.2 \pm 5.6$) compared to the energy spectrum at the time used in this study ($\delta_{nVF} = 2.6$). Additionally, the background emission dominated up to $\approx 50 \, \rm{keV}$. Thus, the later time (02:07:55 UT) was studied here.

In order to fit the energy spectra with \textit{f\_thick\_warm}, additional parameters from imaging observations are required \citep[see the methodology outlined in ][]{2019ApJ...871..225K}, such as the coronal loop length $L$ and hot plasma density $n$ which were determined from imaging, see \S \ref{Observing: Imaging Diagnostics}. The large uncertainties associated with the imaging parameters had various affects on the resulting non-thermal electron properties such as the power law spectral index. Firstly, $L$ (where $L = 16\pm 4$, $18^{+3}_{-4}$, and $24^{+1}_{-14}$ for Flares 1, 2, and 3, respectively) did not have a significant affect, with $\delta_{nVF}$ ranging from $1.8-2.0$ for Flare 1, remaining the same ($\delta_{nVF} = 2.6$) for Flare 2, and ranging from $\delta_{nVF} = 3.0-3.1$ for Flare 3, across the different spectral fitting models.

Coronal plasma density was not able to be obtained using spectral line density diagnostics as there was no Hinode EIS data available for these flares. Instead, the coronal plasma density $n$ was determined in two different ways. Firstly, using Equation \ref{eq:density_emisison_measure}, where EM is determined from spectroscopy and $V$ is determined from imaging (Equation \ref{eq:volume}); the values of $n \, \pm$ uncertainties calculated using this method are given in the fifth column of Table \ref{chap4:tab:imaging_params}. Secondly, the plasma density was also obtained with X-ray spectroscopy alone using the $f\_thick\_warm$ model, where $n$ was not fixed. For Flare 1, spectroscopy gave a value of $n = 7 \times 10^{10} \pm 10.53 \, \rm{cm}^{-3}$, due to the large error this only the imaging value was considered for Flare 1. An average of the two values from these different methods yielded the final coronal plasma density (Table \ref{Table_plasma_properties}) used in the modeling part of this study.

\begin{figure}[t!]
   \centering
    \includegraphics[width = 0.4\linewidth]{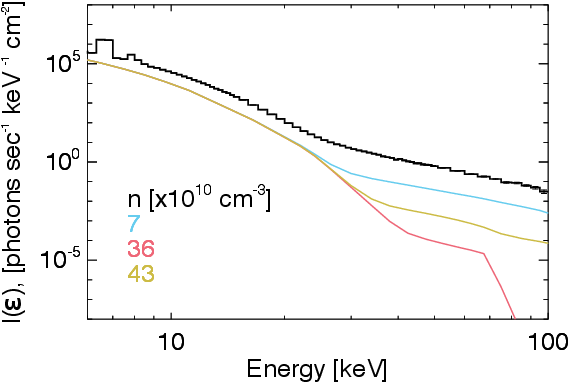}
   \caption{Observed photon flux spectrum of Flare 2 (black) compared to the simulated photon flux spectrum (blue) for simulations using $ n = 7\times 10^{10}, 36\times 10^{10}$,  and $43\times 10^{10} \, \rm{cm}^{-3}$, shown in the left, middle, and right panels, respectively.}
    \label{fig:may_density_comparasion}
\end{figure}

For Flare 2 the $f\_thick\_warm$ spectroscopy method could not constrain the density, producing a value with a very large error ($n = 77 \pm 54 \times 10^{10} \, \rm{cm}^{-3}$). Thus for this flare, only the density produced by imaging will be considered, in which $n = 36^{+7}_{-8}\times 10^{10} \, \rm{cm}^{-3}$. This flare was previously studied by \cite{2019ApJ...871..225K}, here they determined the coronal density to be $\approx 9 \pm 2 \times 10^{10} \, \rm{cm}^{-3}$, lower than in this study. In \cite{2019ApJ...871..225K} the density was calculated using the coronal volume produced by the VIS\_FWDFIT imaging algorithm. They found the volume to be $0.86 \pm  0.20 \times 10^{27} \, \rm{cm}^3 $, which is larger than produced in this study (for Flare 2 $V = 0.28^{+0.09}_{-0.10} \times 10^{27} \, \rm{cm}^3$), and thus, the coronal density is less (see Equation \ref{eq:density_emisison_measure}). As previously discussed, here, the shape of the coronal X-ray source produced by the VIS\_FWDFIT imaging algorithm was larger than the coronal X-ray sources produced by the other imaging algorithms used (i.e. CLEAN, MEM\_NJIT, and EM) and as result was not used. As shown in Figure \ref{fig:may_density_comparasion}, simulations were performed for several densities ($n = [7,36,43] \times 10^{10} \, \rm{cm}^{-3}$, shown in blue, yellow, and pink, respectively) to briefly study how changing density affects the observed photon spectrum. The lowest density used in the simulation was $n = 7 \times 10^{10} \, \rm{cm}^{-3}$ which is the lower density limit found in \cite{2019ApJ...871..225K}. Similarly, the highest density used in the simulation was  $n = 43 \times 10^{10} \, \rm{cm}^{-3}$, to represent the upper density limit determined from imaging in this study. The shape of the simulated photon spectrum best matched the observed photon spectrum at the lower coronal densities\footnote{There is a difference in the observed and simulated photon flux spectra, this could be due to differences in plasma properties or possibly due to the assumption in OSPEX that the electron pitch-angle distribution is angle-integrated (there could be photon spectrum differences due to beaming). However, investigating this fully is beyond the scope of the current study.} studied, i.e., $n = 7 \times 10^{10} \, \rm{cm}^{-3}$; this density will be used for all simulations in this study. 

During a solar flare, the energy released by magnetic reconnection accelerates electrons to energies $> 50$~keV with large fractions of the energy going into non-thermal particles. For each flare studied we also estimate the total non-thermal electron power $P$, [erg s$^{-1}$] from \citep{holman2011implications, 2019ApJ...871..225K} using 
\begin{equation}
    P = \left ( \frac{\delta - 1}{\delta -2 } \right ) \dot{N} E_c
\end{equation}
where $E_c$ is the low energy cut off. All values were obtained from spectroscopy using the $f\_vth+f\_thick\_warm$ model. For Flare 1, 2, and 3 $P = [5.1, \,4.2, \,3.2]\times 10^{27} \, \rm{erg}\,\rm{s}^{-1}$, respectively. This study does not further explore the total non-thermal electron power $P$, however, future studies may use this diagnostics as another useful parameter for building a picture of the acceleration environment in the corona.

\subsection{Imaging Spectroscopy}\label{sect:chap3:imaging_spect}

As previously discussed, X-ray imaging is able to determine the coronal source FWHM and X-ray spectroscopy is used to determine the acceleration timescale. However, in \cite{2023ApJ...946...53S} several useful diagnostics require the use of imaging spectroscopy, such as the ratio of X-ray emission in the chromospheric footpoints, $\eta_{FP}$, spatially resolved spectral indices in the looptop $\delta_{nVF}^{LT}$ and footpoint $\delta_{nVF}^{FP}$, as well as the ratio and difference of these spectral indices. 

\begin{figure*}[t!]
    \centering
    \includegraphics[scale = 0.5]{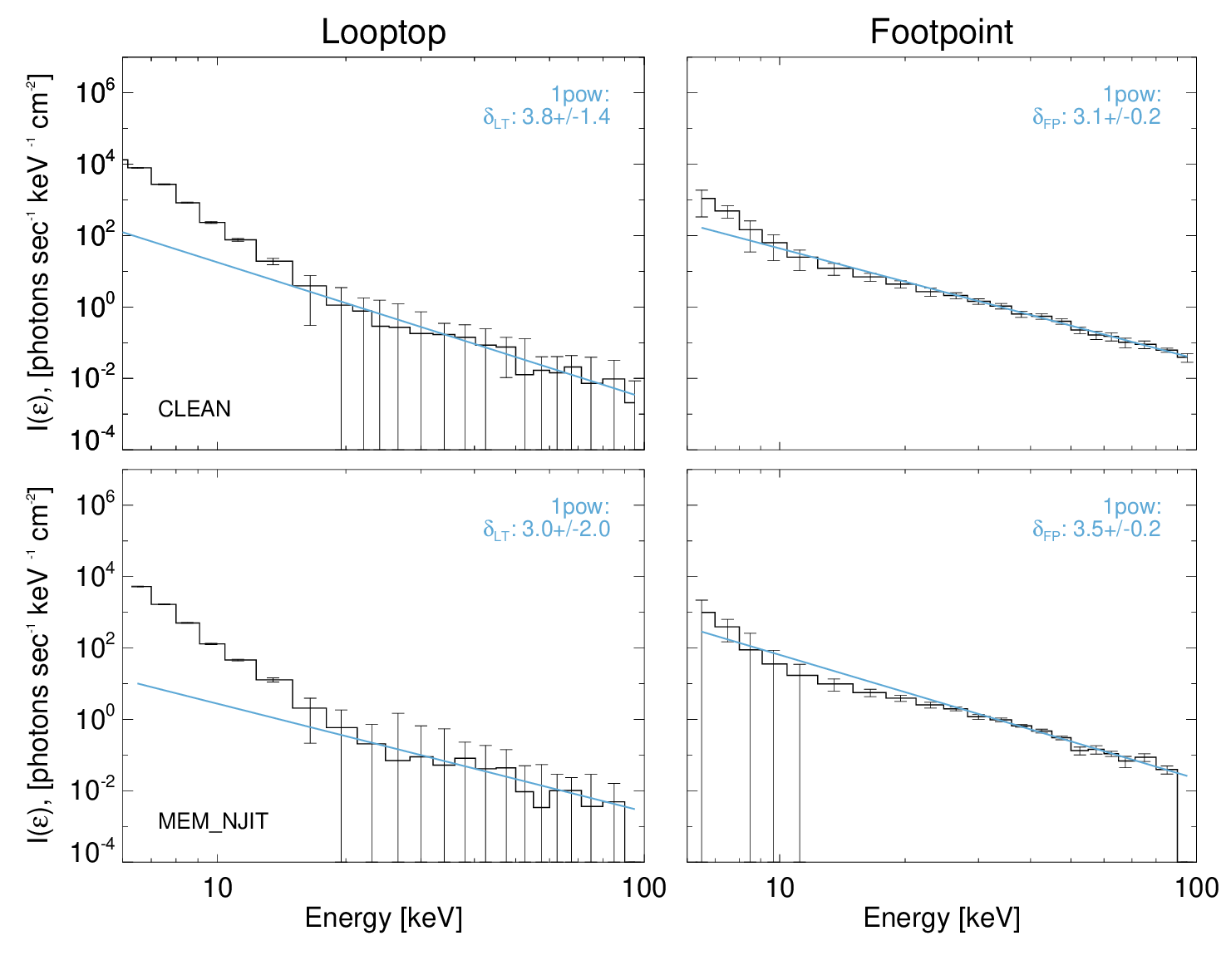}
    \caption{Spatially resolved photon energy spectra for Flare 1 in the coronal looptop (\textit{left column}) and the chromospheric footpoint (\textit{right column}), created using the CLEAN (\textit{top row}) and MEM\_NJIT (\textit{bottom row}) imaging algorithms.}
    \label{fig:feb_d_plot}
\end{figure*}

Spatially resolved energy spectra were created for the coronal looptop source and the chromospheric footpoint sources (defined in \S \ref{Observing: Imaging Diagnostics}). X-ray energy spectra were created for these regions using the relevant imaging algorithms for each flare. For each flare, the spatially-resolved photon spectra were fitted with a power law, i.e., \textit{f\_1pow}, to reduce assumptions about electron transport and plasma properties and to allow for easy comparison between the coronal and footpoint regions. For example, Figure \ref{fig:feb_d_plot} shows the spatially-resolved spectra in the coronal looptop (left) and footpoint (right) sources using the CLEAN (top row) and MEM\_NJIT (bottom row) imaging algorithms, for Flare 1. 
Similarly, the spatially resolved energy spectra for Flare 3 was created using the the MEM\_GE imaging algorithm. In the footpoint region a power law was reliably fitted to non-thermal energies up to $\approx 100 \, \rm{keV}$ for Flares 1 and 2, and up to $\approx 56 \, \rm{keV}$ for Flare 3. For all three flares large uncertainties for energies $> 20 \, \rm{keV}$ make the looptop region futile for this study; as a result, the spectral index in the looptop is not included in this study. Unfortunately, this does reduce the spectral and imaging diagnostics discussed in \cite{2023ApJ...946...53S} that can be used here; the difference ($\delta_{nVF}^{LT} - \delta_{nVF}^{FP}$) and ratio ($\delta_{nVF}^{LT} / \delta_{nVF}^{FP}$) of these spectral indices will not be used.

\section{Kinetic Model}\label{sect:kinetic}

This paper utilizes the kinetic acceleration and transport model described in \cite{2023ApJ...946...53S}; it includes collisional energy losses and scattering as well as turbulent acceleration (see \S \ref{sect:turb_acc}) and turbulent pitch angle scattering (see \S \ref{sect:turb_scat}). Model outputs are discussed in \S \ref{sect:model_outputs}. 

In this study, the low energy cutoff and hence the electron are constrained using the methodology of the warm-target model \citep{2019ApJ...871..225K} which removes the problem with the low energy cutoff and hence provides a more constrained non-thermal electron power. 

\subsection{Turbulent acceleration}\label{sect:turb_acc}
This model considers an extended region of coronal loop turbulence using a spatially dependent turbulent acceleration diffusion coefficient, $D(v,z)$ \citep{2018A&A...612A..64S}, which is a function of both the electron velocity $v$ [cm s$^{-1}$] and the field-aligned coordinate $z$ [cm],
\begin{equation}\label{eq:D_vz}
    D(v,z) = \frac{v^2_{th}}{\uptau_{acc}}\left(\frac{v}{v_{th}}\right)^\alpha  H(z).
\end{equation}
where $v_{th}$ is the thermal velocity, given by $\sqrt{k_B T/m_e}$. 
Equation \ref{eq:D_vz} includes the acceleration timescale $\uptau_{acc}$ [s] and a velocity dependence controlled by the power index $\alpha$. In Equation \ref{eq:D_vz} the spatial distribution of turbulence is described by $H(z)$. As in \cite{2023ApJ...946...53S}, $H(z)$ is chosen to be either decreasing linearly from the loop apex (\textit{linear}),
\begin{equation}
H(z)=  \left (1 - \frac{|z|}{z_B} \right )
\label{Dvz_L}
\end{equation}
or distributed randomly in the loop (\textit{random}) \footnote{Linear and random distributions of turbulence were both observed by \citep{2021ApJ...923...40S} using EUV spectral lines.},
\begin{equation}
H(z)= U[0,1]
\label{Dvz_R}
\end{equation}
or described as a Gaussian centered at the loop apex (\textit{Gaussian}),
\begin{equation}
    H(z) = \text{exp}\left(-\frac{z^2}{2\sigma ^2}\right) \, .
\end{equation}

\subsection{Turbulent pitch-angle scattering}\label{sect:turb_scat}
Similar to \cite{2023ApJ...946...53S}, this study models spatially-dependent turbulent pitch angle scattering using an isotropic pitch angle diffusion coefficient \citep[e.g., ][]{1989ApJ...336..243S,2020A&A...642A..79J}, 
\begin{equation}
D_{\mu\mu}(\mu,v,z)=v\frac{(1-\mu^{2})}{2\lambda_{Ts}} \, H(z) \, ,
\label{eq:turb_scat}
\end{equation}
where $\mu$ is the cosine of the electron pitch-angle to the guiding magnetic field and $\lambda_{Ts}$ [cm] is the turbulent scattering mean free path length.
The properties of turbulent scattering remain largely unconstrained in solar flares. As such, we model two extreme cases, which together to cover a large range of possible underlying scattering and acceleration mechanisms. Firstly, we consider acceleration regions `without short-timescale turbulent scattering' in which $\lambda_{Ts} \rightarrow \infty$. In this case the energy diffusion acts a timescale approximately equal to the scattering mechanism. Secondly, we consider acceleration regions `with short timescale turbulent scattering'. In this case the energy diffusion acts over a timescale greater than the scattering mechanism. For this case, the scattering mean free path length is given by
\begin{equation}
    \lambda_{Ts} =  2 \times 10^8 \, \rm{[cm]}\left(\frac{E}{25\,\text{keV}}\right) 
\end{equation}
as determined in \cite{musset2018diffusive}. 

\subsection{Model outputs}\label{sect:model_outputs}
The model outputs the density-weighted electron spectrum $nVF$ [electrons $\rm{s}^{-1}$ $\rm{cm}^{-1}$ $\rm{keV}^{-1}$] \citep{2003ApJ...595L.115B}, which can be compared directly to the observational X-ray data without assumptions about transport modeling. 
Here, each event is modeled using the observed electron and plasma properties (values given in Table \ref{Table_plasma_properties}): plasma temperature $T$ [K], plasma density $n$ [cm$^{-3}$], (half) loop length from apex to footpoint $L$ [Mm], emission measure EM [cm$^{-3}$], and electron rate $\dot{N}$ [electrons s$^{-1}$].
To constrain the acceleration region properties for each flare, the model outputs are compared to several of the observed spectral and imaging diagnostics discussed in \cite{2023ApJ...946...53S} that were shown to change with acceleration region properties: (1) the size of the coronal looptop source (coronal source FWHM), (2) spectral indices related to the emitting electron energy power law (full flare and spatially resolved), and (3) the ratio of low to high energy emission in the chromospheric footpoint HXR sources.

To obtain equivalent values of the photon spectral index (see \S \ref{sect:chap3:imaging_spect} from the model outputs, the electron spectrum $nVF$ is converted to the photon spectra, $I$, using 
\begin{equation}
I(\epsilon)=\frac{1}{4\pi R^{2}}\int_{E=\epsilon}^{\infty}nVF(E)\,Q(E,\epsilon)dE
\label{eq:photon_spectra}
\end{equation}
where $\epsilon$ is the photon energy, $R$ is the Sun-observation distance and $Q$ is the angle-integrated bremsstrahlung cross section of \cite[][formula 3BN]{1959RvMP...31..920K}.

\section{Results}\label{sect:results}

To constrain the acceleration region properties, we obtain several X-ray spectral and imaging diagnostics (i.e., the coronal source FWHM, the electron energy spectral index, the spatially-resolved photon spectral index in the footpoints, and the ratio of low-to-high energy X-ray emission in the chromosphere) using X-ray observations (see \S \ref{sect:obs}).

The measured spectral and imaging diagnostics may also be obtained by the kinetic model described in \cite{2023ApJ...946...53S}, model outputs are discussed in \S \ref{sect:model_outputs}. Using this model and the observed plasma properties, we simulate the three flares. Multiple simulations were performed for each flare, changing the properties of the acceleration region (Equation \ref{eq:D_vz}) until the values of several spectral and imaging diagnostics produced by the model match the observed values for a given flare.
Table \ref{Table_plasma_properties} summarizes the acceleration region properties which best match the spectral and imaging diagnostics of each flare, as well as the observed plasma properties\footnote{The plasma density given in Table \ref{Table_plasma_properties} is obtained from imaging. \S \ref{sect:spectro_diagnostics} discusses the value of the density used in the kinetic model for each simulation}. Subsequent subsections outline how the X-ray spectral and imaging diagnostics were used to connect simulation and observations and constrain the acceleration region properties

\begin{table*}[t!]
\caption{(\textit{Top}) Observed Plasma properties (see \S \ref{sect:obs}) used in simulation for each flare: Temperature $T$, number density  $n$, coronal loop length from loop apex to footpoint $L$, emission measure EM, and the total non-thermal electron flux $\dot{N}$. Note, the density for Flare 2 has no associated uncertainty, a full explanation is given in \S \ref{sect:spectro_diagnostics}. (\textit{Middle}) The observed spectral and imaging diagnostics: coronal source  FWHM, spatially integrated spectral index $\delta_{nVF}$, spatially resolved X-ray spectral index in the footpoint $\gamma^{FP}$, the ratio of low to high energy emission in the chromosphere $\eta_{FP}^{\text{Xray}}$, and non-thermal electron power $P$. (\textit{Bottom}) The determined acceleration region properties of each flare: the spatial extent $\sigma$, spatial function $H(z)$, the acceleration timescale $\uptau_{acc}$, the velocity dependence $\alpha$, and the presence (yes or no) of short-timescale turbulent scattering.}
\begin{tabular}{cccccc|}
\cline{2-6}
\multirow{3}{*}{} & \multicolumn{5}{|c|}{Plasma Properties} \\ \cline{2-6} 
 \multicolumn{1}{c|}{ } & $T$ & $n$ & $L$ & EM & $\dot{N}$  \\
 \multicolumn{1}{c|}{ } & {[}MK{]} & {[}$\times10^{10} \, \rm{cm}^{-3}${]} & {[}Mm{]} & {[}$\times 10^{49}\, \rm{cm}^{-3}${]} & [$\times 10^{35} \, \rm{electrons}\, {s}^{-1}$] \\ \hline
\multicolumn{1}{|c|}{Flare 1} &  $19\pm 1$ & $6 \pm 1$ & $16\pm4$& $0.10\pm 0.02$ & $0.82 \pm 0.03$  \\
\multicolumn{1}{|c|}{Flare 2} &  $23 \pm 1$ & 7 & $18^{+3}_{-4}$ & $3.7 \pm 0.3$ & $0.51 \pm 0.03$  \\
\multicolumn{1}{|c|}{Flare 3} &  $20 \pm 1$&$6\pm 1$& $24^{+1}_{-14}$ & $0.10 \pm 0.02$ & $1.03 \pm 0.09$\\ \hline
\\ \cline{2-6} 
\multirow{3}{*}{} & \multicolumn{5}{|c|}{Observed Diagnostics} \\ \cline{2-6} 
\multicolumn{1}{c|}{ } & FWHM & $\delta_{nVF}$  & $\gamma_{FP}$ & $\langle \eta_{FP}^{X-ray} \rangle $ & P \\ 
\multicolumn{1}{c|}{ } & {[}Mm{]} &  &  &  & [$\times 10^{27}$ ergs s$^{-1}$] \\ \hline
\multicolumn{1}{|c|}{Flare 1} &  $15.4^{+0.8}_{-0.3}$ & $1.84\pm 0.03$ & $3.3 \pm 0.1$ & $7.4\pm 1.1$ & $5.1 \pm 0.2$ \\
\multicolumn{1}{|c|}{Flare 2} & $14 \pm 1$ & $2.6\pm 0.1 $ & $3.7 \pm 0.4$ & $9.4 \pm 11.2$ & $4.2 \pm 0.3$\\
\multicolumn{1}{|c|}{Flare 3} & $15\pm1$ &$3.0\pm 0.1$& $6.4 \pm 0.6$ & $1.4\pm3.1 $ &$3.2 \pm 0.3$ \\ \hline
\\ \cline{2-6} 

\multirow{3}{*}{} & \multicolumn{5}{|c|}{Acceleration Region Properties} \\ \cline{2-6} 
\multicolumn{1}{c|}{ } & $\sigma$  & $\uptau_{acc}$ & $H(z)$& $\alpha$ & Turbulent Scattering  \\ 
\multicolumn{1}{c|}{ } & {[}Mm{]} & [s] &  &  & yes/no \\\hline
\multicolumn{1}{|c|}{Flare 1} &  $5.4^{+1.4}_{1.0}$ & 7.0 & $l$ & 3 & yes  \\
\multicolumn{1}{|c|}{Flare 2} &  $4.4^{+1.5}_{-1.8}$& 22.0 & $g$  & 3 & yes \\
\multicolumn{1}{|c|}{Flare 3} & $5.5_{-1.5}^{+0.7}$ & 18.4 & $l$ &3 & yes \\ \hline

\end{tabular}

\label{Table_plasma_properties}
\end{table*}

\subsection{Constraining the size of the acceleration region}\label{sect:spatial_extent}
The first step to constraining the properties of the acceleration region is to determine the coronal source FWHM. \cite{2023ApJ...946...53S} showed that the coronal source FWHM increases as the spatial extent of the acceleration region, $\sigma$, increases. The size of the acceleration region is smaller than the measured X-ray coronal source. Furthermore, the size of the coronal source FWHM was found to be independent of the changes in the other acceleration region properties (defined in \S \ref{sect:turb_acc}), i.e., $\uptau_{acc}$, $\alpha$, $H(z)$ and the timescale of turbulent scattering (defined in \S \ref{sect:turb_scat}). In this study, the observed coronal source FWHM was determined for each flare (see \S \ref{Observing: Imaging Diagnostics} for a detailed discussion on how this diagnostic was obtained) and then, in the kinetic model, the spatial extent of the acceleration region was varied until the simulated coronal source FWHM matched the observed coronal source FWHM over the observed energy range of $6-12 \, \rm{keV}$ (for both observations and simulations).

\begin{figure*}[h!]
    \centering
    \includegraphics[width = 0.86\textwidth]{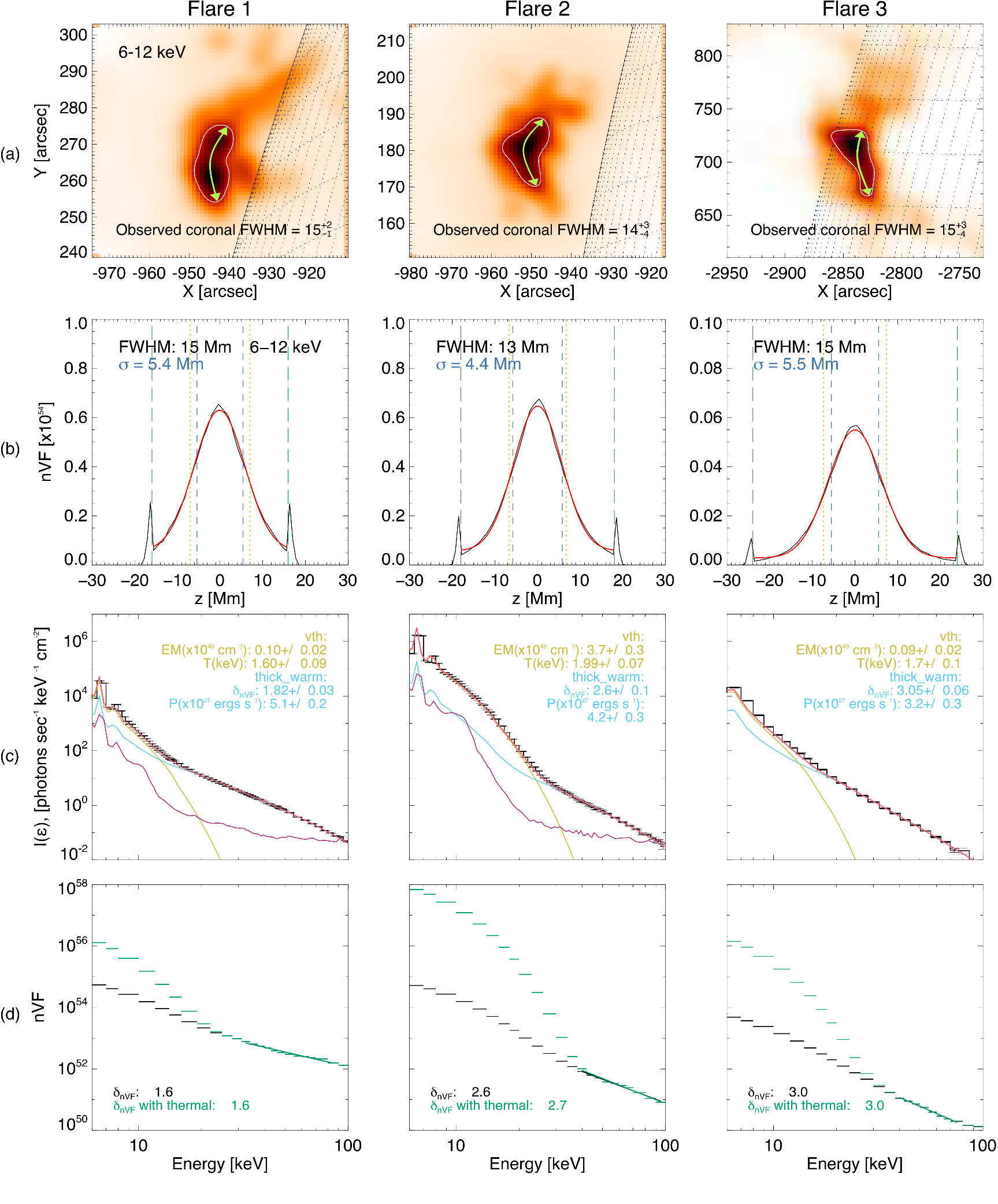}
    \caption{Columns show Flares 1, 2, and 3. Row (a): X-ray image of the flare, in the $6-12 \, \rm{keV}$ energy bin, white contours show the 50\% intensity level. Green arrows depicts the coronal source FWHM. Row (b): Simulated $nVF$ [electrons $\rm{s}^{-1}$ $\rm{cm}^{-2}$ $\rm{keV}^{-1}$] distribution in space $z$, where the loop apex is $z = 0 \, \rm{Mm}$. Green dashed lines indicate the chromospheric boundary. Yellow dotted lines show the observed coronal source FWHM. Blue dashed lines shows the spatial extent of the acceleration region. Coronal emission is fitted with a Gaussian distribution (red). Row (c): Observed spatially-integrated photon flux spectrum (black), integrated over an  $\approx 100 \, \rm{s}$ time period. Fitted with a thermal distribution (yellow) and a warm-thick target model (red). Background emission is shown in purple. Row (d): Simulated electron spatial distribution $nVF$ [electrons $\rm{s}^{-1}$ $\rm{cm}^{-1}$ $\rm{keV}^{-1}$] distribution in energy, with (green) and without (black) the addition of a background thermal component. Solid lines show the spectral fitting (power law).}
    \label{fig:big_fig}    
\end{figure*}

For each of the flares in this study, the observed coronal source FWHM (see Table \ref{Table_plasma_properties}) was found to be $15.4^{+0.8}_{-0.3} \, \rm{Mm}$ (Flare 1), $14\pm1 \, \rm{Mm}$ (Flare 2), and $15\pm1 \, \rm{Mm}$ (Flare 3) . Row (a) of Figure \ref{fig:big_fig} shows $6-12 \, \rm{keV}$ images of Flares 1, 2, and 3 in which the green arrow highlights the observed coronal source FWHM determined from the $50\%$ contour shown in white. The X-ray images in this figure were created using the CLEAN \citep{1974A&AS...15..417H,2002SoPh..210...61H} imaging algorithm for Flares 1 and 2, and the MEM\_GE \citep{2007SoPh..240..241S} algorithm for Flare 3 (see \S \ref{Observing: Imaging Diagnostics} for details). 

In comparison, the simulated coronal source FWHM for each flare was determined by fitting a Gaussian to the spatial distribution of the $6-12 \, \rm{keV}$ emission (the same energy range used to obtain the observed coronal source FWHM) in the corona using
\begin{equation}
    \text{FWHM} = 2\sqrt{2\text{ln}2} \,  \Delta z \, ,
\end{equation}
where $\Delta z$ [Mm] is the standard deviation of the Gaussian distribution. 
For each flare studied, the simulated (density weighted) electron spatial distribution $nVF$ [electrons $\rm{s}^{-1}$ $\rm{cm}^{-1}$ $\rm{keV}^{-1}$] (see \S \ref{sect:model_outputs}) are shown in black in row (b) of Figure \ref{fig:big_fig} and the Gaussian fit is shown in red. The coronal loop apex is at $z = 0 \, \rm{Mm}$ and the chromospheric boundary is shown by the green dashed lines. The coronal source FWHM is highlighted by the yellow dotted lines. The spatial extent $\sigma$ of the acceleration region required to create a simulated coronal source FWHM matching the observed value is given in Table \ref{Table_plasma_properties}, for each flare studied. Additionally, the blue dashed lines in Figure \ref{fig:big_fig} show the spatial extents of the acceleration regions required for each flare. 

In this study the modeling suggests that the spatial extent of the acceleration region covered $28\%$, $24\%$, and $23\%$ of the coronal loop, for Flares 1, 2, and 3, respectively. Thus, for all three events studied, the acceleration region appears to extend down from the loop apex into the loop legs.  This results in a spatial extent $\pm$ uncertainties of $\sigma = 5.4^{+1.4}_{-1.0}\, \rm{Mm}$, $4.4^{+1.5}_{-1.8}\, \rm{Mm}$, and $5.5_{-1.5}^{+0.7}$, for Flares 1, 2, and 3, respectively, where the uncertainty on the simulated value of $\sigma$ was determined by matching the simulation to the observed coronal source FWHM $\pm$ uncertainties. A full description of the spatial extent for different spatial distributions of turbulence (i.e., linear, random, or Gaussian; discussed in \S \ref{sect:spatial_func}) is given in \cite{2023ApJ...946...53S}.

\subsection{Constraining the acceleration timescale}\label{sect:acc_timescale}
Next, the acceleration timescale $\uptau_{acc}$ is constrained, using the spectral index $\delta_{nVF}$ of the emitting electron energy spectra $nVF$.
\cite{2023ApJ...946...53S,2018A&A...612A..64S} showed that changes to the acceleration timescale produced a significant change in the spectral index, such that, as the acceleration timescale increases, the spectral index increases, due to fewer higher energy electrons, producing a softer energy spectrum. 
In this study, after determining the spatial extent of the acceleration region, the observed full flare electron spectral index was determined and then, in the kinetic model, the acceleration timescale of the acceleration region was varied until the simulated spatially integrated electron spectral index matched the observation.

Row (c) of Figure \ref{fig:big_fig} shows full flare X-ray spectra, fitted using isothermal ($f\_vth$) and warm target models ($f\_thick\_warm$), shown in yellow and blue, respectively. Spectral fitting is discussed in \S \ref{sect:spectro_diagnostics}. The spatially integrated photon flux spectrum is shown in black and the full model ($f\_vth+f\_thick\_warm$) is shown in red. For Flares 1 and 2 (observed by RHESSI) the background photon flux spectra is given in purple. The background photon flux spectra for Flare 3 (observed by STIX) has been removed, see \S \ref{sect:spectro_diagnostics}. From the spectral fitting, the observed values of $\delta_{nVF} = 1.84\pm0.03$, $2.6\pm 0.1$, and $3.0 \pm 0.1$ are determined for Flares 1, 2, and 3, respectively (see Table \ref{Table_plasma_properties}).

Simulated $nVF$ energy spectra are shown in row (d) of Figure \ref{fig:big_fig}, for each flare. These energy spectra were created using the individual observed flare properties, i.e., temperature $T$, density $n$, emission measure EM, and and total electron flux $\dot{N}$. The black line shows the energy spectra created by the modeled electrons undergoing acceleration, collisions (including thermalization) and scattering, and the green line shows the same energy spectra but with the addition of a background thermal component, given by 
\begin{equation}
    nVF_{th} = \text{EM} \sqrt{\frac{8}{\pi m_e}} \frac{E}{(k_BT)^{3/2}}e^{-E/k_BT}.
    \label{eq:thermal_component}
\end{equation}
where $E$ [keV] is electron energy, $m_e$ [g] is electron mass and $k_B$ is the Boltzmann constant. For each flare, the isothermal component ($f\_vth$) provides a value of EM and $T$ which is used to compute an additional background thermal component. The solid lines in row (d) of Figure \ref{fig:big_fig} show a power law fit to the simulated $nVF$ spectra, fitted between $40-100 \, \rm{keV}$ for Flares 1 and 2, giving $\delta_{nVF}$. Due to the lack of high energy emission in Flare 3, the simulated spectrum is fitted between $40-74 \, \rm{keV}$. The model was able to successfully obtain energy spectra and spectral indices similar to the observed for all three flares.

As shown in Figure \ref{fig:big_fig} row (d), the addition of a background thermal component did not have a significant effect on $\delta_{nVF}$, with no change to $\delta_{nVF}$ for Flares 1 and 3. There was a slight change to $\delta_{nVF}$ for Flare 2, such that the spectral index increased from $\delta_{nVF} = 2.6$ to $2.7$ with the addition of the background thermal component. Since the inclusion of the background thermal component provides a better comparison to the observed spectra, this study will consider the value of $\delta_{nVF}$ with a background thermal component. 

The simulations which produced outputs that best matched the observations have values of $\uptau_{acc} = 7.0\, \rm{s}$, $22.0 \, \rm{s}$,  and $18.4 \, \rm{s}$, for Flares 1, 2, and 3, respectively.

In the kinetic model the acceleration timescale was very sensitive, slightly altering the value of $\uptau_{acc}$ could produce a spectral index which no longer matched the observation. For example, for Flare 3, increasing the acceleration timescale by $ 0.7 \, \rm{s}$ changed the spectral index from $3.0$ to $4.3$.  Yet,  it should be noted that for the simulations in this study all values of $\uptau_{acc}$ obtained are within the same order of magnitude. Additionally, unlike the acceleration region $\sigma$, $\uptau_{acc}$ is sensitive to variations in other acceleration region properties not fully constrained here. For example, increasing $\alpha$, which controls the velocity dependence (this study only considered $\alpha = 3$; see \S \ref{sect:vel_depend}), produces more non-thermal electrons. Thus, to produce the same photon energy spectrum a lower acceleration timescale is required. Further, although $\uptau_{acc}$ is currently constant (energy independent) in our model, we should also keep in mind that certain underlying mechanisms (e.g., fluctuating magnetic mirror force associated with fast mode turbulence) will produce an acceleration timescale that is energy dependent \citep{2012ApJ...752....4B}. 


\begin{table*}[t!]
\centering
\caption{Observed spectral and imaging diagnostics for each flare (repeated from Table \ref{Table_plasma_properties}), and the acceleration region properties and simulated spectral and imaging diagnostics for different acceleration regions. For each flare model, outputs are shown when the spatial distribution of turbulence is linear ($l$), random ($r$) or Gaussian ($g$). Additionally, results are shown for the presence (yes or no) of short-timescale turbulent scattering (Ts). Acceleration regions which produce spectral and imaging diagnostics which closely match the observation are shown in italics for each flare, with the closest match shown in bold.}
\label{chap4:tab:spatial_func}
\begin{tabular}{cccccccccc}
\cline{7-10}
 &  &  &  &  & \multicolumn{1}{c|}{} & \multicolumn{4}{c|}{Observed Diagnostics} \\ \cline{7-10} 
\multicolumn{1}{l}{} & \multicolumn{1}{l}{} & \multicolumn{1}{l}{} & \multicolumn{1}{l}{} & \multicolumn{1}{l}{} & \multicolumn{1}{l|}{} & FWHM & $\delta_{nVF}$ & $\gamma_{FP}$ & \multicolumn{1}{c|}{$\langle \eta_{FP}^{Xray}\rangle$} \\
 &  &  &  &  & \multicolumn{1}{c|}{} & {[}Mm{]} &  &  & \multicolumn{1}{c|}{} \\ \cline{6-10} 
 &  &  &  & \multicolumn{1}{c|}{} & \multicolumn{1}{c|}{Flare 1} & $15.4^{+0.8}_{-0.3}$ & $1.84\pm0.03$ & $3.3\pm0.1$ & \multicolumn{1}{c|}{$7.4\pm1.1$} \\
\multicolumn{1}{l}{} & \multicolumn{1}{l}{} & \multicolumn{1}{l}{} & \multicolumn{1}{l}{} & \multicolumn{1}{l|}{} & \multicolumn{1}{c|}{Flare 2} & $14\pm1$ & $2.6\pm0.1$ & $3.7\pm0.4$ & \multicolumn{1}{c|}{$9.4\pm11.2$} \\
\multicolumn{1}{l}{} & \multicolumn{1}{l}{} & \multicolumn{1}{l}{} & \multicolumn{1}{l}{} & \multicolumn{1}{l|}{} & \multicolumn{1}{c|}{Flare 3} & $15\pm1$ & $3.0\pm0.1$ & $6.4\pm0.6$ & \multicolumn{1}{c|}{$1.4\pm3.1$} \\ \cline{6-10} 
 &  &  &  &  &  &  &  &  &  \\ \cline{2-10} 
\multicolumn{1}{c|}{} & \multicolumn{5}{c|}{Acceleration Region Properties} & \multicolumn{4}{c|}{Spectral and imaging diagnostics} \\ \cline{2-10} 
\multicolumn{1}{c|}{} & H(z) & Ts & $\sigma$ & $\alpha$ & \multicolumn{1}{c|}{$\uptau_{acc}$} & FWHM & $\delta_{nVF}$ & $\gamma_{FP}$ & \multicolumn{1}{c|}{$\eta_{FP}^{Xray}$} \\
\multicolumn{1}{c|}{} &  &  & {[}Mm{]} &  & \multicolumn{1}{c|}{{[}s{]}} & {[}Mm{]} &  &  & \multicolumn{1}{c|}{} \\ \hline
\multicolumn{1}{|c|}{\multirow{6}{*}{Flare 1}} & $l$ & No & 5.4 & 3 & \multicolumn{1}{c|}{8.7} & 14.5 & 1.9 & 4.0 & \multicolumn{1}{c|}{11.6} \\
\multicolumn {1}{|c|}{} & \textbf{$l$} & \textbf{Yes} & \textbf{5.4} & \textbf{3} & \multicolumn{1}{c|}{\textbf{7.0}} & \textbf{14.0} & \textbf{1.6} & \textbf{3.1} & \multicolumn{1}{c|}{\textbf{6.3}} \\
\multicolumn{1}{|c|}{} & $r$ & No & 5.4 & 3 & \multicolumn{1}{c|}{7.8} & 14.7 & 1.8 & - & \multicolumn{1}{c|}{48.5} \\
\multicolumn{1}{|c|}{} & \textit{$r$} & \textit{Yes} & \textit{5.4} & \textit{3} & \multicolumn{1}{c|}{\textit{9.1}} & \textit{13.9} & \textit{1.7} & \textit{2.9} & \multicolumn{1}{c|}{\textit{5.4}} \\
\multicolumn{1}{|c|}{} & $g$ & No & 5.4 & 3 & \multicolumn{1}{c|}{19.5} & 15.3 & 1.5 & 3.0 & \multicolumn{1}{c|}{5.9} \\
\multicolumn{1}{|c|}{} & \textit{$g$} & \textit{Yes} & \textit{5.4} & \textit{3} & \multicolumn{1}{c|}{\textit{18.2}} & \textit{14.1} & \textit{2.0} & \textit{2.9} & \multicolumn{1}{c|}{\textit{5.8}} \\ \hline
\multicolumn{1}{|c|}{\multirow{6}{*}{Flare 2}} & $l$ & No & 4.4 & 3 & \multicolumn{1}{c|}{12.7} & 13.5 & 2.4 & 3.8 & \multicolumn{1}{c|}{14.1} \\
\multicolumn{1}{|c|}{} & $l$ & Yes & 4.4 & 3 & \multicolumn{1}{c|}{12.1} & 13.8 & 2.4 & 5.4 & \multicolumn{1}{c|}{18.7} \\
\multicolumn{1}{|c|}{} & \textit{$r$} & \textit{No} & \textit{4.4} & \textit{3} & \multicolumn{1}{c|}{\textit{13.9}} & \textit{13.8} & \textit{2.8} & \textit{3.1} & \multicolumn{1}{c|}{\textit{3.1}} \\
\multicolumn{1}{|c|}{} & $r$ & Yes & 4.4 & 3 & \multicolumn{1}{c|}{10.4} & 13.6 & 2.8 & 4.4 & \multicolumn{1}{c|}{35.2} \\
\multicolumn{1}{|c|}{} & $g$ & No & 4.4 & 3 & \multicolumn{1}{c|}{20.2} & 13.7 & 2.3 & 3.0 & \multicolumn{1}{c|}{7.7} \\
\multicolumn{1}{|c|}{} & \textbf{$g$} & \textbf{Yes} & \textbf{4.4} & \textbf{3} & \multicolumn{1}{c|}{\textbf{22.0}} & \textbf{13.3} & \textbf{2.7} & \textbf{3.1} & \multicolumn{1}{c|}{\textbf{7.7}} \\ \hline
\multicolumn{1}{|c|}{\multirow{6}{*}{Flare 3}} & $l$ & No & 5.5 & 3 & \multicolumn{1}{c|}{17.2} & 15.3 & 2.8 & 3.4 & \multicolumn{1}{c|}{10.2} \\
\multicolumn{1}{|c|}{} & \textbf{$l$} & \textbf{Yes} & \textbf{5.5} & \textbf{3} & \multicolumn{1}{c|}{\textbf{18.4}} & \textbf{14.8} & \textbf{3.0} & \textbf{3.2} & \multicolumn{1}{c|}{\textbf{9.9}} \\
\multicolumn{1}{|c|}{} & \textit{$r$} & \textit{No} & \textit{5.5} & \textit{3} & \multicolumn{1}{c|}{\textit{14.3}} & \textit{15.2} & \textit{3.0} & \textit{2.6} & \multicolumn{1}{c|}{\textit{6.9}} \\
\multicolumn{1}{|c|}{} & $r$ & Yes & 5.5 & 3 & \multicolumn{1}{c|}{15.5} & 15.0 & 2.8 & 3.5 & \multicolumn{1}{c|}{11.7} \\
\multicolumn{1}{|c|}{} & $g$ & No & 5.5 & 3 & \multicolumn{1}{c|}{27.5} & 15.8 & 2.8 & - & \multicolumn{1}{c|}{27.0} \\
\multicolumn{1}{|c|}{} & $g$ & Yes & 5.5 & 3 & \multicolumn{1}{c|}{38.8} & 15.0 & 3.2 & - & \multicolumn{1}{c|}{452.1} \\ \hline
\end{tabular}
\end{table*}

\subsection{Constrained acceleration region properties}

\begin{figure*}[t!]
    \centering
    \includegraphics[width = \linewidth]{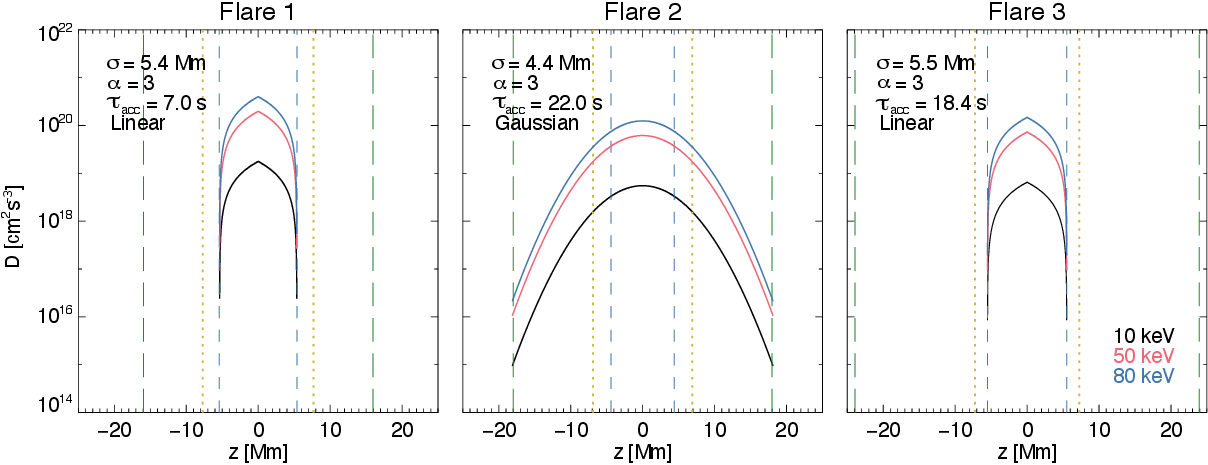}
    \caption{Acceleration diffusion coefficient $D(v,z)$ distribution in space, from the loop apex to chromospheric boundary, using acceleration region properties which produce outputs that best match observations (bold in Table \ref{chap4:tab:spatial_func}) for three different energies, i.e. $10 \, \rm{keV}$ in black, $50 \, \rm{keV}$ in pink, and $80 \, \rm{keV}$ in blue.}
    \label{fig:d_plot}
\end{figure*}

We are unable to fully constrain the remaining acceleration region properties ($H(z)$, $\alpha$ and the turbulent scattering timescale). This is expected as the remaining acceleration region properties require more spectral and imaging diagnostics than was able to be obtained during this study (see Appendix \ref{sect:appdendix1} for additional diagnostics studied). For each flare, we simulated six different acceleration regions, the properties of which are given in Table \ref{chap4:tab:spatial_func}). Here, the regions where multiple diagnostics were within the observational uncertainties (or, as close as possible), are shown in italics. The simulated region which produced the closest match to multiple observations is considered the `best' match and is highlighted in bold in Table \ref{chap4:tab:spatial_func} (Appendix \S \ref{sect:spatial_func} discusses this in detail). For each flare, the `best' match acceleration regions, and the resulting turbulence in the loop via the diffusion coefficient $D(v,z)$, are shown in Figure \ref{fig:d_plot}.

Here, all three acceleration regions include short-timescale turbulent scattering $\uptau_{ts}$ (see \S \ref{sect:turb_scat}, that is, where $\uptau_{ts}$ occurs on timescales much shorter than the collisional time. The acceleration regions for Flares 1 (left panel) and 3 (right panel) look similar; they both have a linearly decreasing distribution of turbulence that disappears after a boundary (i.e., $\sigma=5.4$~Mm and $\sigma=5.5$~Mm). However, for these models, the acceleration timescales are quite different, $\uptau_{acc} =7 \, \rm{s}$ and $\uptau_{acc} =18.4 \, \rm{s}$, respectively. The acceleration timescale of Flare 3 is closer to that of Flare 2 (middle panel), with $\uptau_{acc} = 22.0 \, \rm{s}$. The turbulent spatial distribution in Flare 2 is best described by a Gaussian centered at the loop apex, with less variation in $D(v,z)$ within $\sigma=4.4$~Mm than for a linearly decreasing distribution, but with possible acceleration throughout the loop (although $D(v,z)$ falls steeply with $z$ outside of $\sigma$). However, as previously mentioned, for all flares it was not possible to fully constrain the spatial distribution of turbulence with alternative solutions highlighted in Table \ref{chap4:tab:spatial_func}.

\section{Discussion}\label{sect:discussion}
Over the last few years the dynamic nature of flares, as well as an abundance of flare observations suggesting the presence of turbulence have favored stochastic-type acceleration mechanisms. The properties and the underlying mechanisms creating these turbulent conditions are unconstrained with several proposed mechanisms e.g., interacting plasma waves, the generation of MHD turbulence and/or instabilities related to fragmented current sheets \citep{2023ApJ...947...67R,1997JGR...10214631M,2012A&A...539A..43K}. The new method outlined here, using available X-ray spectral and imaging diagnostics only \citep{2023ApJ...946...53S}, has helped to constrain properties of a coronal looptop acceleration region, for three flares, by comparing HXR observations to kinetic modeling which includes an extended turbulent acceleration region. Our study provides the first steps highlighting the need to use multiple diagnostics and modeling in tandem for this purpose.

Firstly, X-ray imaging diagnostics provided an estimate of the X-ray coronal source FWHM which was used to constrain the spatial extent $\sigma$ of the turbulent acceleration region. For all flares studied, the spatial extent of the turbulent acceleration region covered at least $\sim 25\%$ of the coronal loop irrespective of the differences in other flare electron and plasma properties. This new result suggests that electrons are energized over a substantial fraction of the flare with the bulk of that acceleration occurring within this $25\%$ region. We also note that variations in other acceleration region properties did not affect this result significantly. This new observational result is supported by e.g., the 3D MHD models of \citep{2016ApJ...833...36F,2023ApJ...947...67R}. Here, magnetic reconnection releases a jet of high-speed plasma down into the Sun. When the jet interacts with the SXR coronal loop a shock forms which in turn leads to the formation of a shear-flow driven Kelvin-Helmholtz instability (KHI) above the coronal loop. The newly formed turbulent swirls may then propagate along the coronal loops as Alfvénic perturbations, extending the turbulent region down into the coronal loop. We note that in this model, the strongest turbulence appears at the loop top apex with an extent covering $\sim1/4$ of the loop, similar to our new result here.

Secondly, X-ray spectroscopic diagnostics were used to determine the spectral index of the emitting electron distribution which was used to constrain the acceleration timescale $\uptau_{acc}$. $\uptau_{acc}$ is ultimately a time scale that determines how fast electrons gain energy and it is a vital observable for restricting possible viable stochastic models \citep{2012ApJ...752....4B}. For example, in our system, an acceleration timescale of $\uptau_{acc}$ indicates that all electrons are accelerated to high energies (i.e., above $100 \, \rm{[keV]}$) out of a thermal population within $\uptau_{acc} \, \rm{[s]}$ if collisional effects are neglected. Here, for all three flares, we determine acceleration timescales ranging between 7 and $22 \, \rm{s}$. In terms of the coronal plasma conditions, the found acceleration times are related to the collisional times in each flare as $\uptau_{acc}=[1600,10800,3000] \uptau_{c}$, respectively. We also note that the acceleration timescales obtained for Flares 1 and 3 agree with values obtained using a different X-ray methodology. Recently, \cite{2024arXiv240800213L} also studied Flares 1 and 3. By fitting a $\kappa$ distribution to the X-ray spectra, they calculate $\uptau_{acc}$. For Flare 1, \cite{2024arXiv240800213L} obtains a similar value of $\uptau_{acc} = 7.33 \, \rm{s}$, compared to $\uptau_{acc} = 7 \, \rm{s}$. Additionally, in private correspondence with the author, regarding Flare 3, they determined $\uptau_{acc} =17.86 \, \rm{s}$, again similar to $\uptau_{acc} = 18.4 \, \rm{s}$ determined here. 

The acceleration region properties constrained above are obtained using a fixed value for the (unconstrained) velocity dependence ($\alpha = 3$; see Appendix \ref{sect:vel_depend} for how changing the velocity dependence affects the electron distribution) and an extreme value of the (unconstrained) turbulent scattering timescale (see Appendix \ref{sect:spatial_func} for additional information on the turbulent scattering timescale). Changing these parameters may alter the acceleration properties constrained here. The acceleration timescale especially can be expected to change when changing these properties. However, these values provide initial constraints on these properties.


Although two vital acceleration region properties have been constrained in this study, importantly, our results demonstrate the need for future instruments with improved X-ray imaging spectroscopy, capable of constraining a wider range of acceleration region properties in solar flares. The resolution of current spacecraft instrumentation is a contributing factor to this as they prevent all of the spectral and imaging diagnostics discussed in \cite{2023ApJ...946...53S} from being measured. For example, improved access to spatially-resolved diagnostics such as spectral indices in the coronal looptop and chromospheric footpoints alone, will provide several additional diagnostics (i.e., $\delta_{nVF}^{LT}$, $\delta_{nVF}^{LT} - \delta_{nVF}^{FP}$, and $\delta_{nVF}^{LT} / \delta_{nVF}^{FP}$) that we could not take full advantage of in this study. Additionally, in this study limb flares were selected to obtain a clear separation between coronal and chromospheric sources and minimize projection-effects , this resulted in two of the flare in this study having only one footpoint source visible to the X-ray instrument. Improved instrumentation will allow for clearer separation between sources, increasing the flares that are suitable for use with this method. This may allow spatially-resolved diagnostics to be obtained for each footpoint region in a flare, highlighting any asymmetry in the footpoints and increasing the reliability of the footpoint contribution.
X-ray imaging spectroscopy of the coronal source and HXR footpoints may be improved by the proposal of direct focusing optics, such as the Focusing Optics X-ray Solar Imager (FOXSI) \citep{2009SPIE.7437E..05K,2016SPIE.9905E..0EG}, with the ability to produce cleaner imaging spectroscopy of different sources in the same field of view, which we have demonstrated is vital for further constraining the properties of the acceleration region.

\appendix

\section{Unconstrained acceleration region properties}\label{sect:appdendix1}

The following sections outline how we attempted to use imaging spectroscopy (chromospheric footpoints only) to constrain the remaining acceleration region properties. However, to fully constrain the acceleration region properties improved X-ray imaging spectroscopy is required. 

\subsection{Constraining the spatial distribution of turbulence and scattering timescale}\label{sect:spatial_func}

\cite{2023ApJ...946...53S} considers an extended region of coronal loop turbulence using a spatially dependent turbulent acceleration diffusion coefficient, $D(v,z)$ (see Equation \ref{eq:D_vz}). In Equation \ref{eq:D_vz} the spatial distribution of turbulence is described by $H(z)$; chosen to be either decreasing linearly from the loop apex (\textit{linear}), distributed randomly in the loop (\textit{random}) \footnote{Linear and random distributions of turbulence were both observed by \citep{2021ApJ...923...40S} using EUV spectral lines.}, or described as a Gaussian centered at the loop apex (\textit{Gaussian}). Here, we consider these three distributions of turbulence with and without short-timescale turbulent scattering (see \ref{sect:turb_scat}). Table \ref{Table_plasma_properties} shows these six simulated acceleration regions and their resulting spectral and imaging diagnostics. Figure \ref{fig:comparasion} shows how the spectral and imaging diagnostics output for each of the six simulated acceleration regions compare to the observed, for each flare. The range of observation values are shown by the solid black lines. Acceleration regions using a linearly decreasing spatial function are shown in pink; a random function in blue; and a Gaussian function in yellow. Acceleration regions without short timescale turbulent scattering are shown by the circles and acceleration regions with short timescale turbulent scattering are shown by crosses. The top panel shows values for the coronal source FWHM, followed by spatially-integrated spectral index ($\delta_{nVF}$) in the second panel. In Figure \ref{fig:comparasion}, some data points overlap and as a result appear missing; see Table \ref{chap4:tab:spatial_func} for all values.

Figure \ref{fig:comparasion} shows that the coronal source FWHM and $\delta_{nVF}$ are not able to constrain the spatial distribution of turbulence in the acceleration region alone, thus, spatially-resolved diagnostics were considered.  
Firstly, X-ray imaging spectroscopy was used to find the spatially-resolved photon spectral indices in the coronal looptop and in the footpoints separately. Unfortunately, the spatially-resolved photon index in the coronal looptop was not available for Flares 1 and 3 (see \S \ref{sect:chap3:imaging_spect}) as the spectra was dominated by a thermal component and had large uncertainties at energies $> 20 \, \rm{keV}$. Hence, this study only used the spatially-resolved photon index in the footpoints, $\gamma_{FP}$. 
The third row of Figure \ref{fig:comparasion} compares $\gamma_{FP}$ for the six acceleration regions. For several simulated acceleration regions $\gamma_{FP}$ could not be fitted, due to a lack of high energy emission in the chromosphere, as a result these values are missing (see Table \ref{chap4:tab:spatial_func}).

For Flare 1, several of the simulated acceleration regions produced values of $\gamma_{FP}$ which were similar to the observed value ($\gamma_{FP} = 3.3\pm 0.1$). Comparing all the studied diagnostics (including the ratio of emission in the chromosphere $\eta_{FP}$ discussed in \S \ref{sect:appdendix1}), the acceleration region using a linearly decreasing distribution of turbulence with short-timescale scattering best matched the observed values consistently. However, the other two spatial distributions: random and Gaussian distributions (both with short-timescale turbulent scattering), also produced close matches.

For Flare 2, all simulated acceleration regions produced enough high energy emission that a value of $\gamma_{FP}$ could be determined, several of which were close to the observed value ($\gamma_{FP} = 3.7 \pm 0.4$). However, when considering $\gamma_{FP}$ alongside the other spectral and imaging diagnostics studied, two acceleration regions produced spectral and imaging diagnostics very similar to the observed values: a random distribution of turbulence without short-timescale turbulent scattering and a Gaussian distribution of turbulence with short-timescale turbulent scattering. Of these two acceleration regions, the Gaussian distribution of turbulence with short-timescale turbulent scattering best matched the observation consistently. 

For Flare 3, four of the simulated acceleration regions produced values of $\gamma_{FP}$, however all values were below the observed value of $\gamma_{FP} = 6.4 \pm 0.6$. Considering the other spectral and imaging diagnostics studied, a linear distribution of turbulence with short-timescale turbulent scattering and a random distribution of turbulence without short-timescale turbulent scattering produce values consistent with the observation \ref{chap4:tab:spatial_func}). The linear distribution of turbulence with short-timescale turbulent scattering produced a value of $\gamma_{FP} = 3.2$, which is marginally closer to the observed value (compared to the random distribution of turbulence without short-timescale turbulent scattering where $\gamma_{FP} = 2.6$) and thus, is considered to the closest match to the observation here. 

Overall, using these diagnostics it is very difficult to constrain how the turbulence varies in space in the acceleration region. For the three spatial distributions studied, we can constrain the acceleration timescale, as discussed in \S \ref{sect:acc_timescale}. However, we find that the acceleration timescale is sensitive to different spatial distributions of turbulence, as shown in Table \ref{chap4:tab:spatial_func}. For Flare 1, the acceleration timescale ranges from $\tau_{acc} = 7.0 \, \rm{s}$ when there is a linear distribution of turbulence with short-timescale turbulent scattering to $\tau_{acc} = 19.5 \, \rm{s}$ when a Gaussian distribution of turbulence without short-timescale turbulent scattering is used. Similarly, Flares 2 and 3 experience a range of $\tau_{acc}$ values of $10.4 - 22.0 \, \rm{s}$ and $14.3-38.8 \, \rm{s}$, respectively, when the spatial distribution of turbulence is changed.  
Thus, to further constrain the acceleration timescale, more information must be known about the spatial distribution of turbulence. 
Studying flares with better spatially-resolved X-ray observations may help with this, discussed in \S \ref{sect:discussion}. 

In \cite{2023ApJ...946...53S} none of the spectral and imaging diagnostics studied clearly separated the different spatial distributions of turbulence. Instead EUV imaging spectroscopy may provide a clearer insight into this acceleration region property, as seen in \cite{2021ApJ...923...40S}. As previously stated, there is no EUV imaging spectroscopy observations for the three flares in this study. However, by comparing all of the studied spectral and imaging diagnostics to the observed values for the different spatial distributions of turbulence (see \S \ref{sect:spatial_func}), we were successful in reducing the number of possible acceleration regions.

\subsection{Constraining the velocity dependence}\label{sect:vel_depend}

Among the examined diagnostics, \cite{2023ApJ...946...53S} studied the ratios of X-ray emission in the chromosphere, $\eta_{FP}$. In \cite{2023ApJ...946...53S} $\eta_{FP}$ is defined as the ratio the density weighted electron spectrum $nVF$ in the chromosphere. In this paper, we study the same ratio, but for direct comparison with the X-ray observation, we use the ratio of the X-ray flux $\eta_{FP}^{^{\rm{Xray}}}$. Thus, for the observations, $\eta_{FP}^{^{\rm{Xray}}}$ is defined as the ratio of emission within the footpoint contour for the following energy bins for Flare 1 and Flare 2, 
\begin{equation}
    \eta_{\rm{FP}}^{\rm{Xray}} = \frac{I(\epsilon=20-30 \;{\rm keV})}{I(\epsilon=50-100 \;\rm{keV})} \, . 
    \label{eq:eta_flare_1_2}
\end{equation}
For Flare 3, due to the lack of emission at energies $> 50 \, \rm{keV}$ the energy bins are changed such that,
\begin{equation}
    \eta_{\rm{FP}}^{\rm{Xray}} = \frac{I(\epsilon=20-30 \;{\rm keV})}{I(\epsilon=40-56 \;\rm{keV})} \, ,
    \label{eq:eta_flare_3}
\end{equation}
where $I(\epsilon)$ is the photon spectrum of photon energy $\epsilon$ (see Equation \ref{eq:photon_spectra}).
The white contours in middle and right panels of Figures \ref{fig:Feb_algorithm_comparasion} - \ref{fig:march_algorithm_comparasion} highlight the footpoint region, and thus the emission used to determine $\eta_{\rm{FP}}^{\rm{Xray}}$ for Flares 1, 2, and 3, respectively.
For each event, the spatially-resolved photon spectrum in the footpoint was used to determine the total photon flux in the energy ranges given in Equation \ref{eq:eta_flare_1_2} or \ref{eq:eta_flare_3} (i.e, $20-30 \, \rm{keV}$ and $50-100 \, \rm{keV}$ for Flares 1 and 2, and $20-30 \, \rm{keV}$ and $0-56 \, \rm{keV}$ for Flare 3) for all relevant algorithms, and an average value of $\eta_{FP}^{\rm{Xray}}$ was determined. For Flare 1, where multiple footpoints are visible, $\langle \eta_{FP}^{\rm{Xray}}\rangle$ is used instead, giving the mean value of the $\eta_{FP}^{^{\rm{Xray}}}$ from the two footpoints, 
\begin{equation}
	\langle \eta_{FP}^{\rm{Xray}}\rangle = \frac{\eta_{FP1}^{\rm{Xray}} +  \eta_{FP2}^{\rm{Xray}}}{2}
\end{equation} 
where $\eta_{FP1}^{\rm{Xray}}$ and $\eta_{FP2}^{\rm{Xray}}$ are the ratios of emission in each observed footpoint. $\eta_{FP}^{\rm{Xray}}$ may be studied directly from the X-ray spectra; the result of which agrees with what is shown here. However, obtaining the ratio with energy dependent imaging removes influence of the looptop emission.

\begin{figure}[h!]
    \centering
    \includegraphics{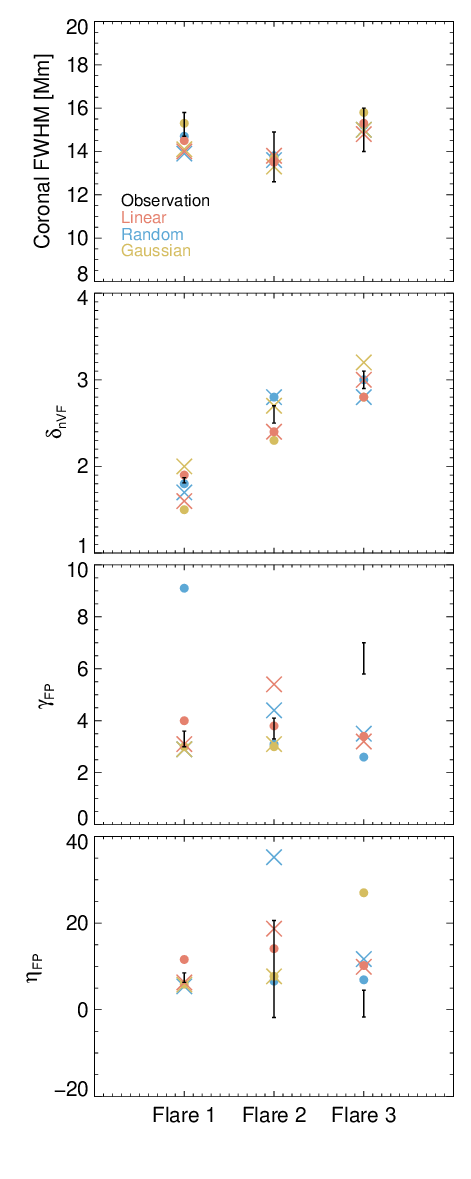}
    \caption{Comparing the observed and modeled spectral and imaging diagnostics: coronal source FWHM, $\delta_{nVF}$, $\gamma_{FP}$, and  $\eta_{FP}^{\text{Xray}}$, from top to bottom, respectively. The acceleration regions for three different spatial functions are shown in different colors: linear spatial functions in pink, random functions in blue, and Gaussian functions in yellow. Crosses and circles show acceleration regions with and without short-timescale turbulent scattering, respectively.}
    \label{fig:comparasion}
\end{figure}

The observed values of $\eta_{FP}^{\text{Xray}}$ are given by the black line in the fourth row of Figure \ref{fig:comparasion}. Flares 1, 2, and 3 had a similar value of $\eta_{FP}^{\text{Xray}}$, $7.4\pm 1.1 $ and $9.4 \pm 11.2$, and $1.4\pm3.1$, respectively. The large uncertainty value on $\eta_{FP}^{\text{Xray}}$ for Flare 2 comes from the large error on the $20-30\, \rm{keV}$ energy bin. 
Although, Flares 2 and 3 in particular have large uncertainties, \cite{2023ApJ...946...53S} showed that changing $\alpha$ by $\pm 1$ can change $\eta_{FP}^{\text{Xray}}$ by several orders of magnitude. Thus, in this initial study, $\eta_{FP}^{\text{Xray}}$ may still be useful for constraining the velocity dependence.

The fourth row in Figure \ref{fig:comparasion} compares the range of the observed values (the black line) and the simulated values (crosses and circles) of $\eta_{FP}^{\text{Xray}}$, which used $\alpha=3$, for each flare. For all three, the simulated value of $\eta_{FP}^{\text{Xray}}$ is of the same order of magnitude as the observation. These values are considered to agree due to the sensitivity of the velocity dependence on $\eta_{FP}^{\text{Xray}}$.

Multiple spectral and imaging diagnostics vary with the velocity dependence and $\uptau_{acc}$. The sensitivity of the velocity dependence, $\alpha$, on the acceleration timescale, $\uptau_{acc}$, was  tested for Flare 1 only, as the simulations and observations produced the closest match, out of the three events studied. As small changes to the velocity dependence greatly affect the spatially-resolved spectral index, simulations using a velocity dependence of $\alpha = 2.5$ and $\alpha = 3.5$ were performed. The acceleration timescale was then changed to produce a spectral index which matched the observed value. This resulted in an acceleration timescale ranging from $\uptau_{acc} = 3.3$ to $28 \, \rm{s}$ when $\alpha =  2.5$ and $3.5$, respectively.  Additionally, once the acceleration timescale for a given value of $\alpha$ matched the simulation, the values of $\eta_{FP}^{\text{Xray}}$ did not significantly vary with $\alpha$ such that, $\eta_{FP}^{\text{Xray}} = 5.7$, $6.3$, and $4.6$ when $\alpha =  2.5$, $3.0$ and $3.5$, respectively. Thus, a value of $\alpha = 3$ is suitable in this study. Additional diagnostics in future studies, summarized in \S \ref{sect:discussion}, may assist in constraining the velocity dependence of turbulence in the acceleration region. 

We stress that several spectral and imaging diagnostics studied in \cite{2023ApJ...946...53S} vary with multiple acceleration region properties. In our study, this became apparent when trying to obtain simulation values for both the spectral index and $\eta_{FP}^{\text{Xray}}$ which matched with observation; there is a balancing act between $\uptau_{acc}$ and $\alpha$. For example, reducing the velocity dependence to increase the value of $\eta_{FP}^{\text{Xray}}$ results in a larger spectral index. To then reduce the spectral index, $\uptau_{acc}$ is altered, which in turn decreased the value of $\eta_{FP}^{\text{Xray}}$. In this study, matching the simulation to the observed spectral index took priority over $\eta_{FP}^{\text{Xray}}$. Spectroscopy diagnostics from RHESSI and STIX are more reliable than the imaging diagnostics. X-ray images are re-created using imagining algorithms whereas the energy spectra comes from the counts on the detectors. As a result, the value of $\uptau_{acc}$ also took priority. Thus, a velocity dependence of $\alpha = 3$ is used, as in \citep{2018A&A...612A..64S}.

\section*{Acknowledgements}
This research was supported by the NASA Living with a Star Jack Eddy Postdoctoral Fellowship Program, administered by UCAR's Cooperative Programs for the
Advancement of Earth System Science (CPAESS) under award $\#$80NSSC22M0097.M
MS gratefully acknowledges financial support from a Northumbria University Research Development Fund (RDF) studentship. MS would like to thank the Royal Astronomical Society for providing a Grant for Study which allowed travel to Graz, Austria to work with Dr. Ewan Dickson on the analysis of STIX data. NLSJ gratefully acknowledges the financial support from the Science and Technology Facilities Council (STFC) Grant ST/V000764/1. NLSJ and JAM gratefully acknowledges financial support from  STFC grant ST/X001008/1. The authors acknowledge IDL support provided by STFC. MS and NLSJ are supported by an international team grant \href{https://teams.issibern.ch/solarflarexray/team/}{“Measuring Solar Flare HXR Directivity using Stereoscopic Observations with SolO/STIX and X-ray Instrumentation at Earth}” from the International Space Sciences Institute (ISSI) Bern, Switzerland. 
For JAM, the research was sponsored by the DynaSun project and has thus received funding under the Horizon Europe programme of the European Union under grant agreement (no. 101131534). Views and opinions expressed are however those of the author(s) only and do not necessarily reflect those of the European Union and therefore the European Union cannot be held responsible for them. This work was also supported by the Engineering and Physical Sciences Research Council (EP/Y037464/1) under the Horizon Europe Guarantee.

\bibliography{bib}{}
\bibliographystyle{aasjournalv7}

\end{document}